\documentclass[12pt]{article}
\usepackage{amsmath,amssymb,mathrsfs,a4wide,comment}
\usepackage{jheppub}
\newcommand{\jtheta}[2]{\genfrac{[}{]}{0pt}{}{#1}{#2}}

\newcommand{\teich}[2]{\raisebox{0.5ex}{$#1$} \scalebox{1.2}{\slash} \raisebox{-0.5ex}{$#2$}}
\newcommand{\moduli}[3]{\raisebox{-0.5ex}{$#1$} \scalebox{1.2}{$\backslash$} \raisebox{0.5ex}{$#2$} \scalebox{1.2}{\slash} \raisebox{-0.5ex}{$#3$}}

\begin{document}

\title{Moduli spaces of non-geometric type II/heterotic dual pairs}
\author{Yoan Gautier, Dan Isra\"el}

\affiliation{Laboratoire de Physique Th\'eorique et Hautes Energies (LPTHE), UMR 7589,
Sorbonne Universit\'e et CNRS, 4 place Jussieu, 75252 Paris Cedex 05, France
}

\emailAdd{gautier@lpthe.jussieu.fr}
\emailAdd{israel@lpthe.jussieu.fr}

\null\vskip10pt
\abstract{
We study the moduli spaces of heterotic/type II dual pairs in four dimensions with $\mathcal{N}=2$ supersymmetry corresponding to non-geometric 
Calabi-Yau backgrounds on the type II side and to T-fold compactifications on the heterotic side. 
The vector multiplets moduli space receives perturbative corrections in the heterotic description only, 
and non-perturbative correction in both descriptions. We derive explicitely the perturbative corrections 
to the heterotic four-dimensional prepotential, using the knowledge of its singularity structure and of the heterotic  perturbative duality group. We 
also derive the exact hypermultiplets moduli space, that receives corrections neither  in the string coupling nor in $\alpha'$. 
}

                              
\maketitle

\section{Introduction}

Non-perturbative dualities between $\mathcal{N}=2$  compactifications to four dimensions play a pivotal role in our understanding 
of string theory dynamics, see~\cite{Aspinwall:2000fd} for a review. A classical example is the duality 
relating heterotic strings compactified on $K3\times T^2$  to type IIA superstrings compactified on a Calabi--Yau three-fold  
that is a $K3$ fibration~\cite{Kachru:1995wm,Ferrara:1995yx}; by applying the duality fiber-wise, as was suggested 
in~\cite{Vafa:1995gm}, this four-dimensional duality is obtained from a more fundamental six-dimensional duality between 
heterotic on $T^4$ and type IIA on a $K3$ surface~\cite{Hull:1994ys}. More general $\mathcal{N}=2$ dualities can be obtained by 
considering, on the heterotic side, quotients of $K3 \times T^2$ by supersymmetry-preserving discrete symmetries 
(that were originally considered in type II~\cite{Aspinwall:1995fw,Ferrara:1989nm,Schwarz:1995bj} 
as duals of CHL compactifications~\cite{Chaudhuri:1995fk}), 
see~\cite{Datta:2015hza} for a recent work.

Recently, we have discovered with Chris Hull a new type of $\mathcal{N}=2$ heterotic/type II  dual pairs in four dimensions~\cite{Gautier:2019qiq}, 
based on previous works~\cite{Israel:2013wwa,Israel:2015efa,Hull:2017llx}. On the type IIA side, 
they can be understood as mirror-folds~\cite{Hull:2004in}, more 
precisely as $K3$ fibrations over a two-torus with transition functions involving {\it mirrored automorphisms}~\cite{Hull:2017llx}, that are 
stringy symmetries combining automorphisms of a K3 surface and of its mirror. On the heterotic side, they could be tought at first glance as 
T-folds consisting in $T^4$ fibrations over a $T^2$ with $O(4,20)$ monodromy twists. However, as was shown in~\cite{Gautier:2019qiq}, perturbative 
consistency of the heterotic model provides a different picture: the $T^4$ fiber has monodromies both around the cycles of the  $T^2$ and around 
the cycles of the T-dual $T^2$. The main goal of the present work is to analyze the moduli spaces of such dual pairs. 

Deriving the quantum moduli space of $\mathcal{N}=2$ four-dimensional compactifications is an essential quantitative test of non-perturbative dualities 
(see {\it e.g.}~\cite{Kachru:1995wm,Kaplunovsky:1995tm,Antoniadis:1995zn,Harvey:1995fq}), 
as quantum corrections on one side are typically mapped to classical expressions on the other side of the duality. 
By supersymmetry, the moduli space of an $\mathcal{N}=2$ compactification splits, at least locally, into the vector multiplets moduli space 
and the hypermultiplets moduli space:
\begin{equation}
\label{eq:moduli_space_factorisation}
\mathcal{M} \cong \mathcal{M}_\textsc{v} \times \mathcal{M}_\textsc{h}\, ,
\end{equation}
where the first factor is a special K\"ahler manifold and the second factor a quaternionic K\"ahler manifold. 
Depending on the duality frame used, each factor may receive $\alpha'$ corrections (if the corresponding factor contains K\"ahler moduli) 
as well as $g_s$ corrections (if the dilaton belongs to one of the corresponding multiplets). 

For standard dualities between type IIA compactified on  Calabi-Yau 3-folds and heterotic on $K3\times T^2$, the quantum 
vector multiplets moduli space has been studied in great detail as, on the type II side, mirror symmetry allows to solve the problem exactly. 
By contrast, the hypermultiplets moduli space is much less understood, as it receives worldsheet instanton corrections on the heterotic side 
and D-brane and NS5-brane instanton corrections on the type IIA side (see~\cite{Alexandrov:2013yva} for a review).

In the models studied in this work, the situation is diffferent as 
the dilaton belongs to a vector multiplet both in the heterotic frame and in the type IIA frame --~unlike what happens for type IIA  
compactifications on Calabi--Yau threefolds. Therefore $\mathcal{M}_\textsc{v}$ may receive one string-loop corrections as well 
as non-perturbative corrections on both sides of the duality and there is no duality frame where the problem can be solved classically. 
We will argue below that the one-loop corrections to the prepotential 
vanish on the type IIA side, and compute the corrections to the prepotential --~hence to the metric on the vector multiplets moduli space~-- 
on the heterotic side, extending the method used in~\cite{Antoniadis:1995ct,deWit:1995dmj}. 

By contrast, the hypermultiplet moduli space $\mathcal{M}_\textsc{h}$ is tree-level exact in both heterotic and type II duality frames. Deriving 
this moduli space on the type IIA side using algebraic geometry tools is not easy as  mirrored automorphisms lack, by definition,  
a geometrical description, see~\cite{Hull:2017llx} for a discussion. Here, using the heterotic description as an asymmetric 
toroidal orbifold, we are able to derive the exact hypermultiplets moduli space (both in $\alpha'$ and $g_s$). 

This work is organized as follows. In section~\ref{sec:review} we review the construction of the non-geometric 
models of~\cite{Gautier:2019qiq} from both type IIA and heterotic viewpoints. In section~\ref{sec:prepot} we discuss 
the general structure of the vector multiplet moduli space as well as the duality groups appearing in the perturbative limits; 
an explicit computation of the one-loop corrections to the prepotential for some of the models is presented in subsection~\ref{subsec:order2_explicit}.
In section~\ref{sec:hyper} we provide a description of the exact hypermultiplets moduli space. Finally, 
conclusions and avenues for future work are given in section~\ref{sec:concl}. Some relevant material about modular forms is provided in the appendices.

\section{Brief presentation of the models}
\label{sec:review}

In this section we briefly summarize the construction of non-geometric Calabi-Yau backgrounds in type IIA 
string theories~\cite{Israel:2013wwa,Israel:2015efa,Hull:2017llx}, 
as well as of their heterotic duals~\cite{Gautier:2019qiq}. 

It is well-known that the moduli space of type IIA compactifications on a K3 surface is given, 
besides the dilaton zero-mode, by the moduli space of non-linear sigma-models with a K3 target space, namely~\cite{Seiberg:1988pf,Aspinwall:1994rg}
\begin{equation}
  \label{eq:NLSMmoduli}
\mathcal{M}_\sigma \cong O(\Gamma_{4,20}) \backslash O(4,20) \slash O(4) \times O(20)\, .
\end{equation}
where the duality group $O(\Gamma_{4,20})$ is the isometry group of  $\Gamma_{4,20}$,  the lattice of total cohomology of the K3 surface 
of signature $(4,20)$. The latter can be decomposed 
as 
\begin{equation}
\Gamma_{4,20} \cong \Gamma_{3,19} \oplus U \, , 
\end{equation}
with $\Gamma_{3,19}$ the second cohomology lattice on K3, and 
$U$ the unique even self-dual 
lattice of signature $(1,1)$. 
Geometrical automorphisms of K3 sigma-models belongs to 
$O(\Gamma_{3,19}) \ltimes \mathbb{Z}_{3,19}$, {\it i.e.} are combinations of large diffeomorphisms and of integral 
shifts of the B-field. 

Of particular interest in the present context are {\it purely non-symplectic automorphims} of order $p$, that 
act on the holomorphic two-form $\omega$ of the K3 surface as $\omega \mapsto \zeta_p\, \omega$, where $\zeta_p$ is 
a primitive $p$-th root of unity. A class of K3 surfaces admitting such automorphisms are hypersurfaces of the form 
\begin{equation}
\label{ref:surf}
z_1^{\, p} + f(z_2,z_3,z_4)=0
\end{equation}
in a weighted projective space, and supersymmetric quotients thereof.  As was shown in~\cite{Hull:2017llx}, using recent mathematical 
results~\cite{mirror0,Comparin:2012ps,mirror1,mirror2}, the diagonal action of an order $p$ purely non-symplectic automorphim $\sigma_p$ acting 
on a surface of type~(\ref{ref:surf}) and of the corresponding order $p$ automorphism $\tilde{\sigma}_p$ of the mirror surface 
(in the Greene--Plesser~\cite{Greene:1990ud}/Berglund--H\"ubsch~\cite{Berglund:1991pp} sense) can be lifted 
to a {\it mirrored automorphism} $\widehat{\sigma}_p$ 
of non-linear sigma models on K3, corresponding to a certain element of the orthogonal group $O(\Gamma_{4,20})$.  
Importantly, the action of  $\widehat{\sigma}_p$ leaves no sub-lattice of $\Gamma_{4,20}$ invariant, and its matrix 
representation $M_p$ can be diagonalized 
over $\mathbb{C}$ as\footnote{Not all 
values of $p$ give consistent mirrored automorphisms as far as we know. One should consider only K3 surfaces realized as 
hypersurfaces in a weighted projective space (for one of the 95 weight systems of Reid and Yonemura) with a 
polynomial of the form~(\ref{ref:surf}). For prime $p$, as discussed in~\cite{Comparin:2012ps},  there exists a 
non-symplectic automorphism only for $p\leqslant 19$, and for $p=11,17$ and $19$ there is no surface of the requested form; hence 
the allowed prime values are $p\in \{ 2,3,5,7,13\}$. For non-prime $p$, 
even in cases where these conditions are met, although mirrors automorphisms exist their matrix form can be 
different from~(\ref{eq:lemma_diag}). We will assume in the following being in one of the favourable cases.} 
\begin{equation}
\label{eq:lemma_diag}
M_p \cong \text{diag}\, \Big(\zeta_p \mathbb{I}_{q},\ldots,\zeta_p^{\, k} \mathbb{I}_{q},\ldots, \zeta_p^{\, p-1} \mathbb{I}_{q}\Big)\, ,
\end{equation}
with $k$ and $p$ coprime and where $q=  24/ \varphi (p)$, $\varphi (p)$ being  Euler's totient function, 
{\it i.e.} the number of positive integers $k\leqslant p$ such that $\text{gcd}(k,p)=1$.

We have introduced in~\cite{Hull:2017llx} type IIA dimensional reductions consisting in $K3$ fibrations 
over a two-torus whose transition functions involve mirrored automorphisms, a  type of generalized 
Scherk-Schwarz reductions with monodromy twists~\cite{Dabholkar:2002sy,Dabholkar:2005ve}. Specifically, for a field $\phi$ of the 
four-dimensional theory transforming in a representation $R$ of the duality group $O(\Gamma_{4,20})$,  consider a reduction 
ansatz  $\phi (x^\mu, y^i) = R[g(y_i)] \phi (x^\mu)$ where $y^i$ are coordinates on $T^2$ and where 
$R[g(y_i)]  \in O(4,20;\mathbb{R})$. We impose further (for each $i=1,2$) that $g(y_i)^{-1}g(y_i+2\pi R_i) \in O(\Gamma_{4,20})$  and 
corresponds to a matrix $M_p$ associated with the action of a mirrored automorphism $\widehat{\sigma}_p$. In the following 
we will consider only models with a monodromy around a single one-cycle of the $T^2$. 

Mirrored automorphisms admit fixed points corresponding to $K3$ Gepner models, or more general $K3$ Landau--Ginzburg 
orbifolds~\cite{Israel:2013wwa,Israel:2015efa}. At one of these fixed points a twisted reduction as above gives a type 
IIA vacuum with $\mathcal{N}=2$ supersymmetry, 
consisting in a freely-acting orbifold of $K3\times T^2$ acting on the $K3$  Landau--Ginzburg orbifold 
as an order $p$ orbifold with a specific discrete torsion 
and on the $T^2$ as an order $p$ shift along a one-cycle. Importantly, the  dilaton lies in a vector 
multiplet (unlike in Calabi--Yau compactifications) and there are no massless Ramond--Ramond forms in the spectrum. 

The heterotic duals of these compactifications were constructed in~\cite{Gautier:2019qiq}. In the heterotic frame the moduli space 
$\mathcal{M}_\sigma$ of eqn.~(\ref{eq:NLSMmoduli}) is the moduli space of compactifications on $T^4$ with arbitrary Wilson lines. The 
 matrix $M_p \in O (\Gamma_{4,20})$ associated with a mirrored $K3$ automorphism is an element of the T-duality group of the heterotic 
Narain lattice of signature $(4,20)$. Starting with heterotic strings on $T^4$ at a point in the toroidal moduli space fixed under the action of $M_p$, 
an $\mathcal{N}=2$ four-dimensional  compactification is obtained as a freely acting toroidal orbifold combining the order $p$ 
twist by  $M_p$ in the $\Gamma_{4,20}$ toroidal lattice with an order $p$ shift along a one-cycle of an extra $T^2$. Importantly, 
the form~(\ref{eq:lemma_diag}) of the matrix $M_p$ (in the appropriate basis) implies that there is no invariant sub-lattice 
of $\Gamma_{4,20}$ under the action of the orbifold, hence no room for non-Abelian gauge symmetry in these models.\footnote{The corresponding type IIA 
statement is that the models have  no BPS D-branes as they don't contain massless Ramond--Ramond ground states.} 

The shift along the $\Gamma_{2,2}$ lattice of the two-torus is characterized by a vector $\delta \in \mathbb{R}^{2,2}$ with 
$\Delta := p\delta \in \Gamma_{2,2}$. As was shown in~\cite{Gautier:2019qiq} modular invariance of the heterotic one-loop partition function, 
hence perturbative consistency of the models, requires that 
\begin{equation}
\label{eq:shift_vect_cond}
\Delta^2 \equiv \left\{ \begin{array}{ll} 2 \mod p &\ , \quad p  \ \text{odd}\\ 2 \mod 2p &\ , \quad  p   \ \text{even}\end{array}\right.
\end{equation}
following the general constraints on asymmetric orbifolds~\cite{Narain:1986qm}. It means that there exists not only an $
O(\Gamma_{4,20})$ monodromy when going 
around a one-cycle of the two-torus but also when going around the T-dual cycle, {\it i.e.} on top of the momentum 
shift there exists a winding shift. 
On the type IIA side of the duality, the heterotic winding charges on $T^2$ become NS5-brane charges so this feature 
is invisible in perturbation theory.\footnote{It 
was already anticipated in~\cite{Vafa:1995gm} that the 'adiabatic argument' used to derive 4d heterotic/type 
II dual pairs from six dimensions allowed in principle 
winding shifts on the heterotic side. 
}

\section{One-loop corrections to the prepotential}
\label{sec:prepot}

In this section we will analyse the space $\mathcal{M}_\mathrm{V}$ spanned by the scalars in the vector multiplets 
(see equation~\eqref{eq:moduli_space_factorisation}). As it has long been known, $\mathcal{N}=2$ supersymmetry imposes that $\mathcal{M}_\mathrm{V}$ 
is a special K\"ahler manifold~\cite{deWit:1984wbb}, whose geometry is completely encoded in a holomorphic function $f$ of the moduli, the prepotential, from which one can derive a K\"ahler metric on $\mathcal{M}_\mathrm{V}$. 

As the axio-dilaton sits in a vector multiplet in both type IIA and heterotic perspectives, the prepotential  (and consequently the K\"ahler metric) 
generically receives corrections from quantum contributions in both cases. It is well known then that, due to the Peccei-Quinn symmetry of the axio-dilaton vector multiplet,  any perturbative correction to $f$ higher than one-loop must vanish as a consequence of $\mathcal{N}=2$ 
supersymmetry~\cite{Antoniadis:1992pm,Harvey:1995fq,Antoniadis:1995ct,deWit:1995dmj,Forger:1997tu}:
\begin{equation}
\label{eq:prepotential_general_form}
  f = f^{(0)} + f^{(1)} + f^{\mathrm{np}},
\end{equation}
$f^{(0)}$, $f^{(1)}$ and $f^{\mathrm{np}}$ being the tree-level, the one-loop and the non-perturbative contributions to the prepotential respectively.

The tree-level contribution to the vector multiplets moduli space is rather easy to understand, as there are generically 
only three vector multiplets for all values of $p$, and is similar to the moduli space of more ordinary $\mathcal{N}=2$ compactifications like 
heterotic strings on $K3 \times T^{\, 2}$ without Wilson lines, see {\it e.g.}~\cite{deWit:1995dmj,Antoniadis:1995ct}. 
One of them contains the axio-dilaton and will be named $S$ in the heterotic description: 
\begin{equation}
S=a + i e^{-\phi}\, ,
\end{equation} 
where the scalar $a$ is the four-dimensional dual of the NS-NS two-form.  The other two are associated with the moduli of the two-torus. The 
moduli parametrise a "Teichm\"uller space" $\mathcal{T}_V$ which may be expressed as the direct product
\begin{equation}
  \label{eq:vector_teich1}
  \mathcal{T}_\textsc{v} = \left(\teich{SL(2)}{U(1)}\right)_S \times \left(\teich{O(2,2)}{O(2) \times O(2)}\right)_{T^2},
\end{equation}
where the $SL(2)/U(1)$ and the $O(2,2)/\left[O(2) \times O(2)\right]$ factors correspond to the axio-dilaton and the two-torus moduli spaces respectively. 
The latter may be further split to give 
\begin{equation}
  \label{eq:vector_teich2}
  \mathcal{T}_\textsc{v} = \left(\teich{SL(2)}{U(1)}\right)_S \times \left(\teich{SL(2)}{U(1)}\right)_T \times \left(\teich{SL(2)}{U(1)}\right)_U.
\end{equation}
with the second $SL(2)/U(1)$ factor (resp. the third) corresponding in the heterotic description 
to the complexified K\"ahler (resp. complex structure) moduli space of the 2-torus, 
respectively $T$ and $U$. This is the Teichm\"uller space of the $STU$ 
model which has already been extensively studied in the literature.  The actual classical moduli space is the quotient of 
this Teichm\"uller space by the discrete duality group that will be described in subsection~\ref{subsec:moduli_dualities}, which 
is a subgroup of the T-duality group~(\ref{eq:two_torus_aut_group}) of two-torus compactifications.

The interesting piece of information accessible to a perturbative study therefore lies in the one-loop correction $f^{(1)}$; as usual with one-loop 
diagrams in string theory, an explicit computation would involve an integration over the worldsheet two-torus complex structure which turns 
out to be hard to handle technically. 

In particular, as was shown in~\cite{Gautier:2019qiq}, for the heterotic models at hand, the integrand of the modular integral does 
not factorise into a product of a Narain lattice and a modular form (neither for $SL(2,\mathbb{Z})$ nor for a congruence subgroup associated 
with the orbifold) because the shift vector has a non-zero norm. As a 
consequence, the powerful procedure developed in~\cite{Angelantonj:2011br,Angelantonj:2012gw,Angelantonj:2013eja,Angelantonj:2015rxa} in order 
to compute one-loop integrals in string theory, based on an expansion of the modular form into Niebur-Poincar\'e series,  cannot be applied 
to our models. Brute force computation could then only result in unappealing quantities not leaving T-duality covariance manifest at best.

Our strategy will therefore be close in essence to the one already used in, {\it e.g.}, \cite{Antoniadis:1995ct,deWit:1995dmj}: the third derivatives of 
the one-loop prepotential $f^{(1)}$ are modular forms in both variables,  and their behaviour under the T-duality group of the orbifolded theory, together 
with the localization of physical singularities related to accidental massless states, 
gives very stringent constraints on them. Using results from modular functions theory and 
physical requirements may then be enough to fix the one-loop correction to the prepotential, 
granting access to all perturbative corrections to the vector multiplets moduli space at once 
while preserving manifest T-duality covariance. We will show in this section how this strategy 
works in general and an explicit result for $f^{(1)}$ in the $p=2$ case will be provided.


\subsection{One-loop correction to the vector multiplet moduli space}

It has long been known that the one-loop correction to the prepotential in $\mathcal{N}=2$ theories is related to the new supersymmetric index 
of~\cite{Cecotti:1992qh}; as shown in~\cite{Antoniadis:1992pm}, the one-loop correction to the K\"ahler potential may be explicitly written as
\begin{equation}
\label{eq:kahler_potential1}
K^{(1)} (T,U) = \frac{i}{16(2 \pi)^3} \int_\mathcal{F} \mathrm{d}^2\mu \, 
\bar\eta^{-2} \mathrm{Tr}_\mathrm{R} \left(J_0 (-1)^{J_0} q^{L_0-\frac{c}{24}} \bar q^{\bar L_0 - \frac{\bar c}{24}}\right)
\end{equation}
with $\int_\mathcal{F} \mathrm{d}^2\mu := \int_\mathcal{F} \frac{\mathrm{d}^2\tau}{\tau_2^2}$ the usual integration  over the $SL(2,\mathbb{Z})$ fundamental 
domain $\mathcal{F}$ of the upper-half plane, $J_0$, $L_0$ and $\bar L_0$ the respective zero-modes of the  $U(1)$ R-current and of the Virasoro 
generators. One can then relate the one-loop prepotential $f^{(1)}$ to the modular integral~(\ref{eq:kahler_potential1}), using (see~\cite{Antoniadis:1995ct}):
\begin{equation}
\label{eq:kahler_potential2}
\partial_T \partial_{\bar T} K^{(1)} = -\frac{i}{8 T_2^2} \left(\partial_T+\frac{i}{T_2}\right) \left(\partial_U + 
\frac{i}{U_2}\right) f^{(1)} + \ \mathrm{h.c}.
\end{equation}
where we have decomposed the heterotic prepotential as:
\begin{equation}
f(S,T,U) = STU + f^{(1)} (T,U) + f^{\mathrm{np}} (S,T,U)\, , 
\end{equation}
respectively the tree-level, one-loop and non-perturbative contributions (the latter being exponentially suppressed in the limit $|S|\to \infty$) following equation~\eqref{eq:prepotential_general_form}.

Under perturbative symetries ({\it i.e.} T-dualities) the one-loop prepotential $f^{(1)}$ does not transform covariantly in general.
For heterotic compactifications with a $T^{\, 2}$ factor, it transforms as a modular function of weight $(-2,-2)$ in $T$ and $U$ 
under $PSL(2,\mathbb{Z})_T \times PSL(2;,\mathbb{Z})_U$ up to order-two polynomials due to monodromies 
around the singularities of the prepotential due to the appearance of additional massless states~\cite{deWit:1995dmj,Antoniadis:1995ct}; 
therefore $f^{(1)}$ may not be expressed in terms of modular forms. In our case the story is similar but, as we will see shortly, the duality group is different.

However, even though the $n$-th derivative of a modular function is generically not modular, the third derivative of a modular function of 
weight -2 turns out to always be a genuine modular function of weight 4; therefore, $\partial_T^3 f^{(1)}$ is a modular function of 
weight $(4,-2)$ in $T$ and $U$ respectively. It turns out that $\partial_T^3 f^{(1)}$ may be directly extracted 
from~\eqref{eq:kahler_potential2} as~\cite{Antoniadis:1995ct}
\begin{equation}
\label{eq:prepotential_kahler}
\partial_T^3 f^{(1)} = -\frac{16 i U_2^2}{T_2^2} \partial_T T_2^2 \partial_T \partial_{\bar U} T_2^2 \partial_T \partial_{\bar T} K^{(1)}.
\end{equation}
In the same way, $\partial_U^3 f^{(1)}$ is a modular function of weight $(-2,4)$ in $T$ and $U$ respectively.

We are then finally ready to extract $\partial_T^3 f^{(1)}$ from~\eqref{eq:kahler_potential1}. As usual in orbifold theories, 
traces must be taken over all (un)twisted sectors and projection onto orbifold-invariant states should be enforced, leading to summing over boundary 
conditions; schematically the one-loop K\"ahler potential~\eqref{eq:kahler_potential1} can be decomposed as:
\begin{equation}
\label{eq:kahler_potential3}
K^{(1)} = \sum_{h,g=0}^{p-1} \int_\mathcal{F} \mathrm{d}^2\mu \, \phi\jtheta{h}{g} \Gamma\jtheta{h}{g}(T,U)
\end{equation}
where $\phi\jtheta{h}{g} (\tau)$ would be, in a standard $K3 \times T^{\, 2}$ compactification without Wilson lines,  a modular form of the congruence subgroup of  $PSL(2,\mathbb{Z})_T \times PSL(2,\mathbb{Z})_U$ associated with the orbifold\footnote{It is not the case for our non-geometric heterotic models, since the contributions from the $K3$ factor and the $T^{\, 2}$ factor are no longer separately modular-invariant, see~\cite{Gautier:2019qiq} for details.} and 
$\Gamma\jtheta{h}{g}$ is the usual sum over the charge lattice of the two-torus defined as
\begin{equation}
\Gamma\jtheta{h}{g} := \sum_{Q \in \Lambda_h} q^{\frac{|Q_L|^2}{2}} \bar q^{\frac{|Q_R|^2}{2}} e^{2 i \pi g (Q,\delta)}
\end{equation}
where we keep the convention from~\cite{Antoniadis:1995ct} for the expression of the left and right charges, namely:
\begin{equation}
\label{eq:charges_def}
Q_L := \frac{\mu_1 \bar U - \mu_2 + \nu_1 \bar T + \nu_2 \bar T \bar U}{\sqrt{2 T_2 U_2}}  \quad , \quad  Q_R := \frac{\mu_1 \bar U - \mu_2 + \nu_1 T + \nu_2 T \bar U}{\sqrt{2 T_2 U_2}} \, ,
\end{equation}
and where 
\begin{equation}(\mu_i,\nu_i) \in \mathbb{Z}^4 + h \delta
\end{equation} 
are the corresponding coordinates of the charges in the sub-lattice $\Lambda_h$ of the Narain lattice associated with the $h$-th twisted sector.

Inserting the worldsheet modular integral~\eqref{eq:kahler_potential3} into the general formula~\eqref{eq:prepotential_kahler} then finally gives:
\begin{equation}
\label{eq:derivative_prepotential_explicit}
\begin{split}
\partial_T^3 f^{(1)} = & \frac{16 i \pi^2 U_2}{T_2^2} \sum_{h,g = 0}^{p-1} \int_\mathcal{F} \mathrm{d}^2 \mu \, \phi\jtheta{h}{g} \\
 & \times \tau_2 \partial_\tau \partial_{\bar\tau} \tau_2^2 \partial_\tau \tau_2^2 \sum_{Q \in \Lambda_h} Q_L \bar Q_R^3 q^{\frac{|Q_L|^2}{2}} \bar q^{\frac{|Q_R|^2}{2}} e^{2 i \pi g (Q,\delta)} \, .
\end{split}
\end{equation}
The above expression should of course be properly renormalised in order to give a well-defined expression 
for $\partial_T^3 f^{(1)}$ (see {\it e.g.}~\cite{Kiritsis:1994ta}); however, it is already useful in the present form in order to determine the location of its poles as well as to understand the duality group of the theory. One may also verify that  $\partial_T^3 f^{(1)}$ behaves as a modular function of weight $(4,-2)$ with respect to $(T,U)$ under a transformation of the $T$-duality group to be derived in the following section from equation~\eqref{eq:derivative_prepotential_explicit} as anticipated.

Obtaining $\partial_T^3 f^{(1)}$ from its modular and analyticity properties requires the knowledge of what happens at large distances in the vector moduli space, 
{\it i.e.} when either $T$ or $U$ tends to a cusp. Any such limit may be understood as a decompactification limit as we will explain below. While 
this is obvious for the $T \rightarrow \infty$ limit,  the cases of the other cusps ($T \rightarrow s$  for $s \in \mathbb{Q}$) 
correspond to a two-torus of vanishing volume with a constant $B$-field background. It is not generically a decompactification limit of the theory of interest {\it per se}, but it is always possible to find another theory for which the corresponding limit is a genuine decompactification limit by acting on $T$ with a $SL(2,\mathbb{Z})$ element. The limits obtained by taking $U$ close to a cusp may be understood in a similar fashion by considering the dual torus instead. As argued in~\cite{Kaplunovsky:1995jw}, it follows then from EFT considerations that one does not expect any pole for  $\partial_T^3 f^{(1)}$  at the cusps.

In the following, we will derive the duality group --~or at least a subgroup thereof~-- of the theory as well as its behavior when one of the moduli gets close to a cusp in both the heterotic and type IIA pictures.

\subsection{Vector multliplets moduli space: dualities}
\label{subsec:moduli_dualities}

As explained above, the actual classical moduli space is given by the quotient of the 
Teichm\"uller space~(\ref{eq:vector_teich1}) by the perturbative duality group acting on the second factor. Deriving 
this duality group, or at least a sufficiently large subgroup thereof, is essential in order to constrain sufficiently the 
modular functions $\partial_T^3 f^{(1)} (T,U)$ and $\partial_U^3 f^{(1)} (T,U)$. After some general remarks we will study first 
the perturbative duality group of the type IIA models, and second of their heterotic duals. 

\subsubsection{Deriving the perturbative duality group}

It is a generic feature of orbifold compactifications to have a duality group different from the parent theory, 
as some symmetries of the latter may not be present in the daughter theory and \textit{vice-versa}. As far as the 
vector multiplet moduli space is concerned, the relevant orbifold action of the models described in section~\ref{sec:review}, 
either in type IIA and in hterotic, is the action  on the two-torus that corresponds to a translation. In the following, we will 
call $\mathcal{G}$ the duality group acting on $\mathcal{M}_\mathrm{V}$.

In the parent heterotic theory, the duality group acting on the torus moduli $T$ and $U$ is given by $O(\Gamma_{2,2})_{T^2}$, $\Gamma_{2,2}$ 
being the charge lattice of the $T^{\, 2}$. A convenient decomposition is:
\begin{equation}
\label{eq:two_torus_aut_group}
  O(\Gamma_{2,2})_{T^2} \cong P \Big[ SL(2,\mathbb{Z})_T \times SL(2,\mathbb{Z})_U \big] \ltimes 
 \big(\mathbb{Z}_2 \times   \mathbb{Z}_2\big)\, .
\end{equation}
In this expression, $P [ SL(2,\mathbb{Z})_T \times SL(2,\mathbb{Z})_U ]$ is the quotient of the group 
$SL(2,\mathbb{Z})_T \times SL(2,\mathbb{Z})_U$ by the involution $(g,h) \mapsto (-g, -h)$, 
while the two $\mathbb{Z}_2$ factors correspond respectively to the exchange\footnote{ 
In the corresponding type IIA duality group the $\mathbb{Z}_2$ factor associated with the $S \leftrightarrow U$ exchange is mirror symmetry on $T^{\, 2}$ and 
maps type IIA to type IIB.}  of $T$ and $U$ and to $(T,U) \mapsto (-\bar T, - \bar U)$.

In general, a shift vector $\delta$ will break $O(\Gamma_{2,2})_{T^2}$ into a smaller subgroup. In order to understand  
the unbroken symmetries of the orbifold models, let us consider a one-loop correction of the schematic form
\begin{equation}
  \label{eq:generic_correlator}
  \left< f(\hat{\mathcal{Q}}) \right>^{(1)}(T,U) = \int_\mathcal{F} \mathrm{d}\mu \sum_{h,g=1}^p \Phi\jtheta{h}{g}(\tau) F\jtheta{h}{g}(\tau;T,U)
\end{equation}
with $f(\hat{\mathcal{Q}})$ depending on the internal charge operators $\hat{\mathcal{Q}}$ of the theory, 
taking values in the lattice~\eqref{eq:charges_def}. A necessary and sufficient condition for a transformation acting on $(T,U)$ to leave~\eqref{eq:generic_correlator} invariant --~and then to be a duality of the theory~-- is that it should mix the sectors 
$(h,g)$ in such a way that the sum over all sectors remains invariant. This is obtained for instance by allowing $(T,U)$ 
to transform as $$F\jtheta{h}{g}(\tau;T,U) \mapsto F\jtheta{h}{g}(\tau;T',U') = F\jtheta{h'}{g'}(\tau;T,U)$$
with $\Phi\jtheta{h'}{g'}(\tau)=\Phi\jtheta{h}{g}(\tau)$.

The function $F\jtheta{h}{g}$ may be explicitly written in terms of a sum over the charge lattice and reads:
\begin{equation}
	\label{eq:F_h_g_generic_form}
  F\jtheta{h}{g} (\tau; T,U) = \sum_{Q \in \Lambda_h} f(Q) e^{-\pi \tau_2 \mathcal{M}^2(Q;T,U) + i \pi \tau \left\langle Q,Q\right\rangle + 2 i \pi g \left\langle Q,\delta\right\rangle}
\end{equation}
where the scalar product $\left\langle\cdot , \cdot\right\rangle$ is defined with respect to $\Gamma_{2,2}$ and where the mass function 
is given by:
\begin{equation}
        \label{eq:mass_function}
        \mathcal{M}^2\left(Q=\left[\begin{pmatrix}\mu_1 \\ \mu_2\end{pmatrix}, \begin{pmatrix}\nu_1 \\ \nu_2\end{pmatrix}\right];T,U\right) := \frac{1}{T_2 U_2} \left|\begin{pmatrix}1 & T \end{pmatrix} \begin{pmatrix} \mu_1 & -\mu_2 \\ \nu_2 & \nu_1 \end{pmatrix} \begin{pmatrix} U \\ 1 \end{pmatrix} \right|^2.
\end{equation}
An arbitrary transformation $\hat{g} \in \mathcal{G}$ acts on $X = T,U$ as 
\begin{equation}
X \mapsto \rho_X(\hat{g}) \cdot X\end{equation} 
with $\rho_X$ some representation of $\mathcal{G}$. It may easily be seen from equation~\eqref{eq:F_h_g_generic_form} that $F\jtheta{h}{g}$ may only transform into $F\jtheta{h'}{g'}$ under the action of $\hat{g}$ if there exists some representation $\rho$ of $\mathcal{G}$ such that 
\begin{equation}
\mathcal{M}^2(Q;\rho_T(\hat{g}) \cdot T,\rho_U(\hat{g}) \cdot U) = \mathcal{M}^2(\rho(\hat{g}) \cdot Q;T,U)
\end{equation} 
for any  $Q \in \Lambda$. Then, setting $(T',U') := (\rho_T(g) \cdot T,\rho_U(g) \cdot U)$ for clarity, $F\jtheta{h}{g}(T,U)$ transforms as:
\begin{equation}
F\jtheta{h}{g} (\tau;T',U')  =\!\!\! \sum_{Q \in \rho(\hat{g}) \cdot \Lambda_h}\!\!\! f(\rho(\hat{g}^{-1}) \cdot Q) e^{-\pi \tau_2 \mathcal{M}^2(Q;T',U') + i \pi \tau \left\langle \rho(\hat{g}^{-1}) \cdot Q,\rho(\hat{g}^{-1}) \cdot Q\right\rangle + 2 i \pi g \left\langle \rho(\hat{g}^{-1}) \cdot Q,\delta\right\rangle} 
\end{equation}

Therefore, a necessary condition for $F\jtheta{h}{g} (\tau;T',U')$ to be identified to $F\jtheta{h'}{g'} (\tau;T,U)$ for some $h'$ and $g'$ is to have 
\begin{equation}
\left\langle \rho(\hat{g}^{-1}) \cdot Q,\rho(\hat{g}^{-1}) \cdot Q\right\rangle = \left\langle Q, Q\right\rangle \qquad \forall Q \in \Lambda_h \, . 
\end{equation}
Assuming furthermore that $\rho(\hat{g})$ acts on the charges $Q$ linearly, this is equivalent to requiring that $\rho(\hat{g})$ belongs to $O(\Gamma_{2,2} \otimes \mathbb{R})$. Imposing this restriction, the above equation now reads:
\begin{equation}
F\jtheta{h}{g} (\tau;T',U')  = \sum_{Q \in \rho(\hat{g}) \cdot \Lambda_h} f(\rho(\hat{g}^{-1})  \cdot Q) e^{-\pi \tau_2 \mathcal{M}^2(Q;T,U) + i \pi \tau \left\langle Q,Q\right\rangle + 2 i \pi g \left\langle Q,\rho(\hat{g}) \cdot \delta\right\rangle} \, .
\end{equation}

The transformation $(T,U) \mapsto (T',U')$ may then be a duality of the theory only if it preserves the full charge lattice, that is if $\rho(\hat{g}) \cdot \Lambda = \Lambda$, and if 
\begin{equation}
f(\rho(\hat{g})^{-1} \cdot Q) = J(\hat{g};T,U) f(Q) \qquad \forall Q \in \Lambda 
\end{equation}
for some function $J$ independent of the charge vector $Q$. The first condition will allow us in the following to identify the duality group of the theory from either type IIA and heterotic points of view while the second one is only reflecting the usual behaviour of modular covariant correlator functions.

\subsubsection{Perturbative type IIA symmetries}
\label{subsec:typeII_duality}

The perturbative duality group acting on  Teichm\"uller space~(\ref{eq:vector_teich2}) will be very different depending on whether one considers the theory in the 
type IIA or in the  heterotic perturbative regime. 
In addition to the exchange of the $T$ and $S$ moduli (that follows from heterotic/type IIA duality in four dimensions), 
the shift vectors $\delta_\textsc{iia}$ and $\delta_\textsc{Het}$ used in the respective perturbative 
limits are of different nature (light-like in the 
former case but not in the latter) as we have reviewed in section~\ref{sec:review}.

 We will start by looking at the type IIA duality frame, where $S$ and $U$ are the two-torus moduli and $T$ the axio-dilaton. 
The model is understood as an orbifold of $K3 \times T^{\, 2}$ acting as an order $p$ mirrored automorphism on the $K3$ factor 
and as a shift along the two-torus. In the type IIA theory, the shift vector satisfies $\delta_\textsc{iia}^{\, 2} = 0$ hence may be chosen as:
\begin{equation}
\label{eq:type_IIA_shift}
\delta_\textsc{iia}=\left(\frac{1}{p},0,0,0\right)\, ,
\end{equation}
{\it i.e.} as an order $p$ momentum shift along one circle. 

Let us first focus on the component of $O(\Gamma_{2,2} \otimes \mathbb{R})$ connected to the identity, which acts on the moduli of the torus as:
\begin{equation}
  (S,U) \mapsto (g_S \cdot S , g_U \cdot U) := \left(\frac{a S + b}{c S + d},\frac{a' U + b'}{c' U + d'}\right)\ ,\ ad-bc=a'd'-b'c'=1.
\end{equation}
The parametrisation given in~\eqref{eq:mass_function} allows one to infer straightforwardly the corresponding action on the charges of the lattice:
\begin{equation}
\label{eq:charge_vector_transform}
\begin{pmatrix} \mu_1 & - \mu_2 \\ \nu_2 & \nu_1 \end{pmatrix} \mapsto \begin{pmatrix} d & b \\ c & a \end{pmatrix} \begin{pmatrix} \mu_1 & - \mu_2 \\ \nu_2 & \nu_1 \end{pmatrix} \begin{pmatrix} a' & b' \\ c' & d' \end{pmatrix}
\end{equation}
with $(\mu_i,\nu_i) \in \mathbb{Z}^4 + h \delta_\textsc{iia}$ in the $h$-th twisted sector. As we have just seen, a necessary 
condition for a transformation to give rise to a duality of the theory is that it must preserve the charge lattice 
$\Lambda$. As a result, $g_S$ and $g_U$ must have the form:
\begin{equation}
g_S = \frac{1}{\sqrt{e}} \begin{pmatrix} a e & b \\ c p & d e \end{pmatrix}\ ,\ g_U = 
\frac{1}{\sqrt{e}} \begin{pmatrix} a' e & b' p \\ c' & d' e \end{pmatrix}\ ,\ e|p\ ,\ \gcd\left(e,\frac{p}{e}\right)=1 \, .
\end{equation}

These are the Atkin-Lehner involutions\footnote{The term ``involution" is related to the fact that the square of an Atkin-Lehner involution is in $\Gamma_0(p)$ and acts therefore trivially on the corresponding modular forms.} already encountered in~\cite{Angelantonj:2015rxa,Persson:2015jka}. Such an action on $S$ and $U$ is not generically a duality of the theory though as a vector $Q$ in the $h$ sector is mapped to another one in the $h'$ sector, with $h' = d a' h e + b c' n_1 \frac{p}{e}$ and $n_1$ the winding number of $Q$ around the first circle of $T^{\, 2}$. An arbitrary transformation then generically splits $\Gamma\jtheta{h}{g}$ into a sum of contributions coming from various sectors, so that deriving the full duality group would require a more in-depth analysis of the details of the model.

For our purposes it will be sufficient to identify only a simpler subgroup of the whole duality group, by restricting to transformations 
which preserve each sub-lattice $\Lambda_h$ separately. This may easily be obtained from the above transformations by setting $e=1$; 
the corresponding transformations all belong to $\mathcal{G}_p^\textsc{iia}$, defined as:
\begin{equation}
\mathcal{G}^\textsc{iia}_p := \left\lbrace (g, g') \in \Gamma_0(p)_S \times \Gamma^0(p)_U \middle| g_{11} = g'_{11}, g_{22} = g'_{22}\ \mathrm{mod}\ p\right\rbrace
\end{equation}
with $\Gamma_0(p)$ (resp. $\Gamma^0(p)$) the group of $SL_2(\mathbb{Z})$-matrices whose lower (resp. upper) 
off-diagonal component vanishes modulo $p$. One has in particular 
$$\Gamma_1(p)_S \times \Gamma^1(p)_U \subsetneq \mathcal{G}_p^\textsc{iia} \subsetneq \Gamma_0(p)_S \times \Gamma^0(p)_U\, $$ 
with the congruence subgroups $\Gamma_1(p) = \left\lbrace g \in \Gamma_0(p) \middle| g_{11}=g_{22}=1\ \mathrm{mod}\ p\right\rbrace$ and, in a similar way, $\Gamma^1(p) = \left\lbrace g \in \Gamma^0(p) \middle| g_{11}=g_{22}=1\ \mathrm{mod}\ p\right\rbrace$.

We now turn to a brief analysis of the behavior of the models at the cusps of $\Gamma_0 (p)$.  First,  the $S\rightarrow i\infty$ limit is the type IIA decompactification limit of the two-torus and, due to the freely-acting nature of the orbifold that acts as a momentum shift along $T^2$, it is 
described by a type IIA theory compactified on $K3$ (see {\it e.g.}~\cite{Kiritsis:1996xd}), thereby effectively restoring $\mathcal{N}=4$ supersymmetry. 

When $S$ gets close to one of the other inequivalent cusps, one may analyse the situation by performing a double T-duality along the two-torus and going to the decompactification limit of this dual torus. Following~\cite{Persson:2015jka}, the type IIA worldsheet theory obtained by a double T-duality can be described as an orbifold ($\mathcal{M}/\langle \hat \sigma_p \rangle \times \tilde{S}^1)/G_p \times  \tilde{S}^1$, where $\mathcal{M}/\langle \hat \sigma_p \rangle$ is the quotient of the $K3$ CFT $\mathcal{M}$ by the mirrored automorphism $\hat \sigma_p$ and $G_p$ is an order $p$ cyclic group acting on the first factor as the quantum symmetry of the orbifold $\mathcal{M}/\langle \hat \sigma_p \rangle$ and on the second factor as an order $p$ shift.

A crucial property of the mirrored automorphisms is that the orbifold $\mathcal{M}/\langle \hat \sigma_p \rangle$ is actually isomorphic to 
the original $K3$ CFT $\mathcal{M}$, owing to fractional mirror symmetry~\cite{Israel:2015efa}. Hence the theory obtained after 
double T-duality is exactly of the same type as the original theory, so that the same conclusions hold for the behavior at all cusps: $\mathcal{N}=4$ is restored.

The analysis of the cusps in the $U$-plane  is similar to what we have obtained for the behavior at the 
cusps in the $S$-plane, considering the mirror type IIB model instead of type IIA.

\subsubsection{Perturbative heterotic symmetries}

We now consider the heterotic dual of the model, {\it i.e.} an orbifold of $T^{\, 4} \times T^{\, 2}$ acting as an automorphism of the 
$\Gamma_{4,20}$ Narain lattice on $T^{\, 4}$ and as a shift along the two-torus whose K\"ahler moduli is $T$ and whose complex structure 
moduli is $U$. We restrict the analysis to the case of an orbifold by a group isomorphic to $\mathbb{Z}_p$; then, as shown 
in~\cite{Gautier:2019qiq}, one may choose the shift vector to have components\footnote{Here, it is understood that the shift 
vector is chosen so that the heterotic theory is dual to the type IIA theory with shift vector defined in~\eqref{eq:type_IIA_shift}.}
\begin{equation}
\label{eq:delta_het}
\delta_\textsc{het} = \left(\frac{1}{p},0,\frac{1}{p},0\right)
\end{equation} 
with no loss of generality, see eqn.~(\ref{eq:shift_vect_cond}). 

The derivation of the perturbative duality group in this case goes along the same lines as in the type IIA case. 
The non-vanishing norm of the shift vector forbids in this case any sector-mixing behavior comparable to what we 
had observed in the type IIA case; to be more precise, a vector in $\Lambda_h$ may only be mapped to a vector 
in $\Lambda_{h'}$ if $h^2=h'^2\ \mathrm{mod}\ p$. In particular, for $p$ prime, this means that $\Lambda_h$ 
may only be mapped to $\Lambda_{\pm h}$. Let's consider a transformation:
\begin{equation}
        (T,U) \mapsto (g_T \cdot T , g_U \cdot U) := \left(\frac{a T + b}{c T + d},\frac{a' U + b'}{c' U + d'}\right)\ ,\ ad-bc=a'd'-b'c'=1\, .
\end{equation}
In order to avoid unnecessary complications, we will restrict from now to the cases where all coefficients in the above equations are integers, the 
rationale being that possibile dualities with non-integer coefficients will not be needed for the analysis of the prepotential  below.

First, one realizes that the two $\mathbb{Z}_2$ factors of~\eqref{eq:two_torus_aut_group} from the mother theory duality group remain symmetries of the daughter theory. Indeed, though preserving one $\mathbb{Z}_2$ was expected as the orbifold leaves a one-cycle of the two-torus invariant, preserving the second one as well is somewhat more unusual. As this T-duality exchanges momentum and winding number, it may remain a symmetry of the orbifold theory only if the shift of the orbifold acts in a similar fashion on both the two-torus and its dual, which is the case with the shift vector~(\ref{eq:delta_het}).

Second, imposing in addition that a duality must preserve the charge lattice $\Lambda$ and keeping in mind that $\Phi\jtheta{h}{g}$ in equation~\eqref{eq:generic_correlator} must be equal to $\Phi\jtheta{-h}{-g}$ as a result of CPT invariance, one can check that the perturbative duality group of the heterotic theory must contain:
\begin{equation}
\label{eq:het_duality_group}
\mathcal{G}^\textsc{het}_p := \left\lbrace (g, g') \in SL_2(\mathbb{Z})_T \times SL_2(\mathbb{Z})_U \middle| g'= \pm \sigma_3  g \sigma_3 \ \mathrm{mod}\ p \right\rbrace \ltimes \big(\mathbb{Z}_2 \times   \mathbb{Z}_2\big).
\end{equation}
as a subgroup, $\sigma_3$ being the third Pauli matrix. Equivalently, one has 
\begin{equation}
\mathcal{G}_p^\textsc{het} = \Big( SL(2,\mathbb{Z}) \times \Gamma(p) \Big) \ltimes \big(\mathbb{Z}_2 \times   \mathbb{Z}_2\big) \, ,
\end{equation}
since the condition $g'=\pm\sigma_3 g \sigma_3 \mod p$ can be solved as $g' = \pm\gamma \sigma_3 g \sigma_3$ with $g \in SL(2,\mathbb{Z})$ and $\gamma \in \Gamma (p)$, $\Gamma(p) :=\left\lbrace g \in SL(2,Z) \middle| g = \mathbb{I} \ \mathrm{mod}\ p\right\rbrace$ being the principal congruence subgroup of level $p$. 

Acting non-trivially on only one of the complex moduli of $T^{\, 2}$ (that is, setting either $g=\mathbb{I}$ or $g'=\mathbb{I}$ in the above definition) gives the subgroup:
\begin{equation}
\label{eq:het_duality_subgroup}
\Gamma(p)_T \times \Gamma(p)_U \ltimes \big(\mathbb{Z}_2 \times   \mathbb{Z}_2\big) \subsetneq \mathcal{G}_p^\textsc{het}\, .
\end{equation}
For the rest of the discussion, we will focus on this subgroup and won't attempt to derive the full heterotic 
perturbative duality group of the theory.

The behavior of the theory when going to large distances in the moduli space may be extracted directly from \textit{e.g.} the partition function of the model in this case along the lines of~\cite{Kiritsis:1996xd} and using its explicit form computed in~\cite{Gautier:2019qiq}. It turns out that when either $T$ or $U$ tend to any cusp of $\Gamma(p)$, the theory may be described by a heterotic string theory on a four-torus, restoring once again $\mathcal{N}=4$ supersymmetries (for the $U$ modulus, this is requested by heterotic/type IIA duality).

It is worth mentioning that the fact that the gravitini masses vanish in those limits, thereby restoring $\mathcal{N}=4$ supersymmetry, 
does not imply in general the vanishing of quantities which would vanish in a ``genuine" $\mathcal{N}=4$ theory;  in particular, it does not imply that the Yukawa coupling $\partial_T^3 f^{(1)}$ tends to zero when $T$ or $U$ tends to a cusp.\footnote{
Imposing those constraints on the modular form $\partial_T^3 f^{(1)}$ would actually be too stringent for most values of $p$.}

As explained in~\cite{Kiritsis:1996xd}, while the mass of the two massive gravitini tend to zero,  some charged states 
may be lighter in this limit.  Those light charged states would always keep 
track of the original $\mathcal{N}=2$ behavior of the theory no matter how small one makes the gravitini mass; 
consequently, there would be no reason to expect, say, $\partial_T^3 f^{(1)}$ to be vanishing in this limit. 

We will now argue that this is the case for the models considered in this work. Let us first consider the string states corresponding to the 
massive gravitini, and the large volume limit $|T|\to \infty$. These states are of the form:
\begin{equation}
|\Psi_\textsc{g}\rangle = \big(|s_0;p^\mu\rangle_\textsc{r} \otimes |s';0 \rangle_\textsc{r} \otimes |\hat{s};P_L\rangle_\textsc{r} \big)  \otimes \big( 
\widetilde{\alpha}^{\mu}_{-1} |p^\mu\rangle  \otimes |0\rangle \otimes  |P_R\rangle\big)\, ,
\end{equation}
where we have chosen for the $T^4$ CFT a Ramond ground state $|s';0 \rangle_\textsc{r}$ with unit charge under the action of the $\mathbb{Z}_p$ orbifold. The 
momentum $(P_L,P_R)$ along the $T^2$ is chosen such that $|\Psi_\textsc{g}\rangle$ is even 
under the orbifold projection associated with the shift vector~(\ref{eq:delta_het}). 
Given the mass formula~(\ref{eq:mass_function}) the lightest such state  has $\mu_1=1$ and $\mu_2 = \nu_1=\nu_2=0$ ({\it i.e.} one unit of momentum along the first circle of the two-torus) and the gravitino mass is given by $M_g = |U|/\sqrt{U_2 T_2}$.   

Light charged states can be obtained easily from the Kaluza-Klein modes of the $T$ and $U$ vector multiplets. Specifically, consider a state of the form 
\begin{equation}
|\Psi_\textsc{k}\rangle = \big(|s_0;p^\mu\rangle_\textsc{r} \otimes |s;0 \rangle_\textsc{r} \otimes |\hat{s};Q_L\rangle_\textsc{r} \big)  \otimes \big( 
 |p^\mu\rangle  \otimes |0\rangle \otimes \widetilde{\alpha}^{1}_{-1} |Q_R\rangle\big)\, ,
\end{equation}
where the Ramond ground state $|s;0 \rangle_\textsc{r}$ of the $T^4$ CFT is neutral under the orbifold action, and where the oscillator $\widetilde{\alpha}^{1}_{-1}$
is along the first circle of the two-torus. The lightest such states that are invariant under the orbifold projection have $\mu_2 =1$ and 
$\mu_1 = \nu_1 = \nu_2 = 0$ ({\it i.e.} one unit of momentum along the second circle of the two-torus) and their mass is given by 
$M_\textsc{k} = 1/\sqrt{U_2 T_2}$.  

Thus $M_\textsc{g}/M_\textsc{k} = |U|$ which is greater than one inside the fundamental domain $\mathcal{F}_0$ of $SL(2,\mathbb{Z})_U$. The other 
parts of the fundamental domain of $\Gamma(p)_U$ are obtained as $g\cdot \mathcal{F}_0$ for some $g \in SL(2,\mathbb{Z})$ and can be analyzed 
along the same lines, by transforming the shift vector accordingly.


Finally, one may wonder  whether the vector multiplets moduli space of the putative non-perturbative $\mathcal{N}=2$ theory has an exact 
duality group, related to the perturbative groups $\mathcal{G}_p^\textsc{het}$ and $ \mathcal{G}_p^\mathrm{IIA}$.  
On general grounds one expects that the heterotic vector multiplets moduli space
gets corrected by NS5-branes instanton effects breaking the perturbative duality group 
(see however~\cite{Ferrara:1995yx} as an exception to this  rule).  
It has been shown for instance that  T-dualities  of heterotic strings on $K3\times T^2$ do not survive quantum effects as 
can be seen from the Calabi--Yau type IIA  dual, where the corresponding worldsheet instanton effects are known thanks to mirror symmetry~\cite{Aspinwall:1999ii}. 
In the present case,  since there is no duality frame in which the vector multiplets 
moduli space is classical,  there is  no obvious way to adress this question.

Dualities acting on the $(SL(2,\mathbb{Z})/U(1))_U$ factor of the space~(\ref{eq:vector_teich2}) alone, given that 
there is no frame in which $U$ is the axio-dilaton, may still be exact symmetries of the quantum theory if they appear on both sides 
of the duality. We have shown above that the IIA perturbative group contains a congruence subgroup 
$\Gamma^1 (p)_U$, while the heterotic perturbative group contains a smaller congruence subgroup $\Gamma(p)_U$ 
of $SL(2,\mathbb{Z})$. 

A duality $g \in \Gamma^1(p)_U \backslash \Gamma(p)_U$ is a symmetry on the type IIA side but 
does not belong to the factorized subgroup~(\ref{eq:het_duality_subgroup}) on the heterotic side. If we consider the larger  
subgroup~(\ref{eq:het_duality_group}) of the heterotic duality group, $g \in \Gamma^1(p)_U \backslash \Gamma(p)_U$ remains a 
symmetry of the theory if accompanied by a non-trivial transformation in $SL(2,\mathbb{Z})_T$. 

From the type IIA side, this could be a problem as $T$ is now the axio-dilaton. However, for any such 
$g =${\tiny $\begin{pmatrix}1&b\\0&1\end{pmatrix}$}$\ \mathrm{mod}\ p$, an appropriate transformation  of $T$
would be given by $g'=${\tiny $\begin{pmatrix}1&b\\0&1\end{pmatrix}$}, {\it i.e.} by an integral shift of the NS-NS axion 
$T \mapsto T + b$, $b \in \mathbb{Z}$. This transformation preserves the perturbative regime $\text{Im}\, (T) \rightarrow \infty$ and 
this discrete Peccey-Quinn symmetry is expected to remain a symmetry of the quantum theory.  

In conclusion, one may speculate that $\Gamma^1 (p)_U$ acting on the vector moduli space is an exact duality of the $\mathcal{N}=2$ quantum theory. 
Other exact dualities symmetries acting on the hypermultiplets moduli space, which does not receive $g_s$ corrections, will 
be given in section~\ref{sec:hyper}.

\subsection{Heterotic case: singularities of the prepotential}
\label{sec:singularities}

Our goal in this subsection is to derive $\partial_T^3 f^{(1)} (T,U)$, which is a modular function of 
weight $(4,-2)$ in $T$ and $U$, using  its singularity structure and its behavior at the cusps, applying theorems of modular forms.

Understanding the location of the singularities is fairly easy from an effective field theory (EFT) point of view. To get 
an effective $\mathcal{N}=2$ supergravity theory from the underlying string theory, one has to integrate all heavy fields; 
there may be points in the vector moduli space  where otherwise massive states become massive, resulting in a breakdown of the original effective field theory. 
As a result, the prepotential becomes singular at such a point leading in a pole of order one in $\partial_T^3 f^{(1)}$~\cite{deWit:1995dmj}. 

From the conformal weights of the operators of the heterotic theory, one learns that the mass of a state satisfies:
\begin{equation}
\frac{m^2}{2} = \frac{|Q_L|^2}{2}+N_L+a_L = \frac{|Q_R|^2}{2}+N_R+a_R
\end{equation}
with $N_L$ ($N_R$) the excitation number and $a_L$ (resp. $a_R$) the zero-point energies of the left- (resp. right-) 
moving fields. $a_L$ and $a_R$ where explicitly computed in~\cite{Gautier:2019qiq} and read:
\begin{equation}
\label{eq:zero_point_energies}
\begin{split}
a_L & = \left\lbrace \begin{array}{l} \frac{h}{p} - \frac{1}{2} \ \mathrm{if}\ h \leq \frac{p}{2} \\\\ -\frac{h}{p} + \frac{1}{2} \ \mathrm{if}\ h \geq \frac{p}{2} \end{array} \right. \\ \\
a_R & = \frac{h^2}{p^2} - \frac{h}{p} - \left(\frac{\gcd(h,p)}{p}\right)^2  \prod_{\substack{q | p \\ q\, \mathrm{prime}}} (-q)
\end{split}
\end{equation}
in the $h$-th twisted sector (the inequalities in the expression of $a_L$ being valid for the representative of $h$ in $\mathbb{Z}_p$ such that $0<h<p$). In the untwisted sector, $a_L = -\frac{1}{2}$ and $a_R = - 1$ as usual; a state with non-vanishing charge may therefore be massless in this sector if and only if
\begin{equation}
\left\lbrace \begin{array}{l} |Q_L|^2 = 0 \\\\ |Q_R|^2 = 2 \end{array} \right. \, .
\end{equation}
While there is no such state in general, equation~\eqref{eq:charges_def} implies that a state of charge $Q=(m_i,n_i) \in \mathbb{Z}^4$ may become massless if
\begin{equation}
\label{eq:prepotential_singularities1}
T=\begin{pmatrix} -m_1 & m_2 \\ n_2 & n_1 \end{pmatrix} \cdot U \ \mathrm{and} \ \begin{pmatrix} -m_1 & m_2 \\ n_2 & n_1 \end{pmatrix} \in SL_2(\mathbb{Z}) \, .
\end{equation}

So far, the situation is the same as in the mother theory; the orbifold projection will furthermore select some allowed charges $Q$. 
At the end of the day, assuming as before that the basis is chosen such that the shift vector has non-vanishing components along 
the first cycle of the two-torus only, see eqn.~(\ref{eq:delta_het}), 
the generically massive states which become massless at some points in the vector moduli space have charges satisfying:
\begin{equation}
\label{eq:prepotential_singularities2}
\begin{pmatrix} -m_1 & m_2 \\ n_2 & n_1 \end{pmatrix} = \begin{pmatrix} a + \epsilon & b \\ c & a \end{pmatrix} \quad \mathrm{mod}\ p\, ,
\end{equation}
with $\epsilon \in \left\lbrace 0, \pm 1 \right \rbrace$. The states satisfying the above equation with $\epsilon=0$  
belong to $\mathcal{N}=2$ vector multiplets and correspond to non-abelian enhancements of the gauge symmetry; 
in contrast, the case $\epsilon = \pm 1$ corresponds to states belonging to charged hypermultiplets, 
then resulting in additional matter states without any enhancement of the gauge group. 
In either case, these states are responsible for the appearance of single poles in $\partial_T^3 f^{(1)}$ at the lines of the moduli space given in~\eqref{eq:prepotential_singularities1}. 

New singular lines absent in the mother theory could also occur if extra charged massless states come from the twisted sectors, 
which may happen only if the zero-point energy of the right-moving fields $a_R$ is negative. It is worthwhile noticing that such 
a state would necessarily belong to a hypermultiplet, as only twisted oscillators of the $T^4$ have a small enough conformal dimension 
to fulfill the massless condition coming from the supersymmetric side of the CFT. As it turns out, even though the analysis of 
the situation goes along the same lines as in the untwisted sector case, 
it may not be performed keeping $p$ (and $h$, the label of the twisted sector) generic.

As usual, finding which states may become massless or not for given values of the moduli $T$ and $U$ may also easily be done 
by computing the new supersymmetric index $\mathcal{I}$ of~\cite{Cecotti:1992qh}, whose worldsheet modular integral 
gives the one-loop K\"ahler potential~(\ref{eq:kahler_potential1}). Defining $\mathcal{I}\jtheta{h}{g}$ 
as the contribution from the $h$-th twisted sector with the insertion of the generator of the orbifold to 
the power $g$, it is easy to show that:
\begin{equation}
	\mathcal{I}\jtheta{0}{g} = -\frac{i}{\bar\eta^6(\tau)} \left[\prod_{d \left| \frac{p}{(g,p)}\right.} \bar\eta(d \tau)^{-\mu\left(\frac{p}{d \times (g,p)}\right)}\right]^{24/\varphi\left(\frac{p}{(g,p)}\right)} \left(\prod_{i=1}^2 \bar\vartheta\jtheta{1}{1+2 g s_i/p}\right) \Gamma\jtheta{0}{g}
\end{equation}
for $g \neq 0$ and $\mathcal{I}\jtheta{0}{0} = 0$, as usual.\footnote{In the above derivation, equation~\eqref{eq:theta_to_eta} has been used in order to derive an expression more suited for numerical computations but equivalent to the more traditional form involving more $\vartheta$-functions.} 
It is quite straightforward to obtain from there any $\mathcal{I}\jtheta{h}{g}$ acting with elements of $SL_2(\mathbb{Z})$ on the above; if $p$ is prime, the contribution $\mathcal{I}_h$ from the $h$-th twisted sector to $\mathcal{I}$ then reads:
\begin{subequations}
	\begin{equation}
		\mathcal{I}_0 = -\frac{i p^{12/(p-1)}}{p \bar\eta^6(\tau)} \left(\frac{\bar\eta(\tau)}{\bar\eta(p \tau)}\right)^{24/(p-1)} \sum_{g = 1}^{p-1} \left(\prod_{i=1}^2 \bar\vartheta\jtheta{1}{1+2 g s_i/p}\right) \Gamma\jtheta{0}{g}
	\end{equation}
	
	\begin{equation}
		\mathcal{I}_{h \neq 0} = \frac{i p^{12/(p-1)}}{p \bar\eta^6(\tau)} \sum_{g = 0}^{p-1} \left(\frac{\bar\eta(\tau + h^{-1} g)}{\bar\eta\left(\frac{\tau + h^{-1}g}{p}\right)}\right)^{24/(p-1)}  \left(\prod_{i=1}^2 \bar\vartheta\jtheta{1+2 h s_i/p}{1+2 g s_i/p}\right) \Gamma\jtheta{h}{g}
	\end{equation}
\end{subequations}
with $h^{-1}$ the inverse of $h$ in $\mathbb{Z}_p^\times$.

Charged massless states give rise to divergences in $\partial_T^3 f^{(1)}$ through the contribution of unphysical tachyons coming from the non-supersymmetric side of the worldsheet CFT; therefore, the knowledge of $\mathcal{I}_h$ allows to look for such tachyons in its expansion around $\tau \rightarrow i\infty$. This way, one may check for instance that no charged states coming from the twisted sector(s) become massless at any point of the $T^{\, 2}$ moduli space for $p=2$. Of course, the new supersymmetric index also gives information about the residues of $\partial_T^3 f^{(1)}$, even though those may also be determined by purely effective field theory considerations. In general, one finds~\cite{deWit:1995dmj}

\begin{equation}
	\label{eq:EFT_residues}
	\underset{U \rightarrow \gamma \cdot T}{\mathrm{Res}} \partial_T^3 f^{(1)} = \frac{\beta_\gamma}{16\pi^2} \frac{\det^2(\gamma)}{J^{4}(\gamma,T)}
\end{equation}
where $\beta_\gamma$ is the beta function coefficient associated to the gauge group under which the corresponding charged massless fields are charged and where $J(\gamma,T) :=c T+d$ for $$\gamma = \begin{pmatrix} a & b\\c & d\end{pmatrix}.$$

\subsection{Vector multiplets moduli space: quantum corrections}

In the following, we show that the above is sufficient to determine a closed form for $\partial_T^3 f^{(1)}$ in terms 
of modular functions of $\Gamma(p)$ for any value of $p$, at least in principle. We then proceed to the explicit 
computation of the corrections for the models with  $p=2$.

\subsubsection*{Type IIA viewpoint}

The above analysis has been mainly focused on the heterotic side of the theory because the derivation of $\partial_T^3 f^{(1)}$ is more involved 
in this case. Indeed, in the perturbative type IIA regime, unlike in heterotic, there is no non-Abelian enhancement of the gauge symmetry in the $(S,U)$ moduli space, hence no associated logarithmic singularities of the gauge couplings. We have checked as well that, at least for $p=2$, there are no charged hypermultiplets from the 
twisted sectors of the asymmetric orbifold  of the Gepner model that could become massless.

As argued in~\cite{Kaplunovsky:1995jw,deWit:1995dmj}, using simple effective field theory considerations, the second derivatives 
of $f^{(1)}_\textsc{iia} (S,U)$, {\it i.e.} the gauge couplings, should grow at most linearly in $S$ and $U$ in the decompactification 
limit hence one does not expect a pole of $f^\textsc{iia} (S,U)$ at the cusps $S= i\infty$ and $U=i\infty$. Moreover, as we have argued in section~\ref{subsec:typeII_duality}, the theory obtained when $S$ or $U$ tend to any other cusp may be equivalently described in terms of
 a theory isomorphic to the original ones, so that the same arguments show that $h^\textsc{iia} (S,U)$ is holomorphic at all cusps.

In conclusion, using the fact that the space of negative weights modular forms is empty for any congruence sugroup, 
the type IIA perturbative corrections should vanish:
\begin{equation}
f_\textsc{iia}^{(1)} (S,U) = 0 \, .
\end{equation}
Therefore, on the type IIA side, all the corrections to the prepotential in~\eqref{eq:prepotential_general_form} are of non-perturbative nature.

\subsubsection*{Heterotic viewpoint}

In the heterotic picture, the one-loop correction to the prepotential $f^{(1)}_{\textsc{het}}$ cannot vanish  as it would be inconsistent 
with the analysis of its poles and of their residues. 

The idea here is to express $\partial_T^3 f_\textsc{het}^{(1)}$ in terms 
of modular functions of $\Gamma(p)$, thereby making duality~\eqref{eq:het_duality_subgroup} manifest. It is a standard fact 
from modular form theory that a modular function $f_k$ of weight $p$ with respect to $\Gamma(p)$ 
satisfies\footnote{The $p=2$ case must be treated separately because $\Gamma(2)$ is the only principal congruence subgroup 
containing $-\mathbb{I}$, leading to a difference from a factor of 2 between the two equations.}~\cite{diamond2006first}
\begin{subequations}
\label{eq:valence_formula}
	\begin{equation}
		\sum_{x \in X(2)} \mathrm{ord}_x(f_k) = \frac{k}{2} \qquad \mathrm{if\ } p = 2
	\end{equation}
	\begin{equation}
		\sum_{x \in X(p)} \mathrm{ord}_x(f_k) = \frac{k}{24} \left|SL_2(\mathbb{Z}):\Gamma(p)\right|\qquad \mathrm{if\ } p>2
	\end{equation}
\end{subequations}
where $X(p)$ is the compactification of the quotient of the upper-half plane $\mathcal{H}$ by $\Gamma(p)$ (that is the space $\mathcal{H}\slash \Gamma(p)$ to which one add the corresponding cusps) and where $\mathrm{ord}_x(f_k)$ is the order of $f_k$ at $x$ in the complex analysis sense, that is the least power in the Laurent expansion of $f_k$ around $x$ with non-vanishing coefficient.

Let us then consider our function $\partial_T^3 f_\textsc{het}^{(1)}(T,U)$ as a function of $U$ with fixed parameter $T$ for a moment. As we have seen, the only poles of $\partial_T^3 f_\textsc{het}^{(1)}$ must be simple and located along singular lines of the form $U = \gamma \cdot T$ in the $(T,U)$ moduli space; we will denote in the following the set of all such $\gamma$'s as $\Gamma_\mathrm{sing}$. This means that the orders of $\partial_T^3 f_\textsc{het}^{(1)}$ --~seen as a function of $U$ only~-- must satisfy:
\begin{equation}
	\left\lbrace\begin{array}{l}
		\mathrm{ord}_{U} \partial_T^3 f_\textsc{het}^{(1)} = -1\ \forall\, U \in \Gamma_\mathrm{sing} \cdot T \\\\
		\mathrm{ord}_{U} \partial_T^3 f_\textsc{het}^{(1)} \geq 0\ \forall\, U \notin \Gamma_\mathrm{sing} \cdot T
	\end{array}\right. .
\end{equation}

Recalling that $\partial_T^3 f_\textsc{het}^{(1)}$ is a modular function of weight $(4,-2)$ under $\Gamma(p)_T \times \Gamma(p)_U$, equation~\eqref{eq:valence_formula} shows that there is no much freedom left by these requirements; to be more specific, there will be exactly 5 free parameters if $p=2$ and $$\left|\Gamma_\mathrm{sing}\right| - \frac{1}{12} \left|SL_2(\mathbb{Z}) : \Gamma(p)\right|$$ free parameters if $p>2$ left unfixed by just imposing the location of the poles of $\partial_T^3 f_\textsc{het}^{(1)}$. Mathematically, these free parameters correspond to the location of the zeroes of $\partial_T^3 f_\textsc{het}^{(1)}$ and are as such harder to fix through physical requirements only in general.

The additional data about the residues of $\partial_T^3 f_\textsc{het}^{(1)}$ give even more information and allows to fix at least some of the above parameters. Focusing on the divisor of $\partial_T^3 f_\textsc{het}^{(1)}$ seen as a function of $U$ fixes it up to an overall multiplicative constant with respect to $U$, that is up to an overall multiplicative function of $T$ only; each residue computation lead to a constraint on the free parameters from above. Fixing the overall multiplicative function of $T$ leaves then $\left|\Gamma_\mathrm{sing}\right|-1$  constraints in total,  which are non-necessarily independent. 

At the end of the day, we have more constraints than free parameters and are as such entitled to hope that these would be enough to completely determine $\partial_T^3 f_\textsc{het}^{(1)}$ for any $p$. If it happened not to be the case for some orders $p$ though, fixing the remaining free parameters would have to be done in a different but not necessarily cumbersome way; indeed, computing only the first few terms in the expansion of $\partial_T^3 f_\textsc{het}^{(1)}$ would be enough to fix it completely. At worst, our analysis would then allow to express an infinite series expansion in terms of somewhat easier to handle modular forms. We now turn to illustrating the above strategy in the $p=2$ case.

\subsection{\texorpdfstring{$p=2$}{p=2} case: explicit derivation}
\label{subsec:order2_explicit}

In the following subsection, we will illustrate the  strategy outlined above by computing explicitely the one-loop corrections to the prepotential for non-geometric 
models based on orbifolds of order $p=2$, from their modular properties.\footnote{This 
particular class of compactifications has previously been considered in the litterature~\cite{Vafa:1995gm}, 
albeit formulated in a different way. Indeed, in the type IIA picture, such orbifold of Gepner model may be understood as 
acting as $(-1)^{F_L}$ (together with a momentum shift along the $T^2$). However, as far as the authors know, no derivation of the corrections 
to the prepotential had been given for this model before.}

The analysis of section~\ref{sec:singularities} show that there are 6 singular lines in the $(T,U)$-moduli space mod $\Gamma(2)_T \times \Gamma(2)_U$; more precisely, along the $T=U,U+1,-1/U$ and $U/(U+1)$ lines, one additional vector multiplet becomes massless, resulting in a $U(1)^2 \rightarrow SU(2) \times U(1)$ gauge symmetry enhancement. Along the $T=-1/(U+1)$ and $(U-1)/U$ lines, two charged hypermultiplets become massless, leading to additional matter content without gauge symmetry being enhanced. 

Accidentally, it turns out that a more compact way of parametrising the singular lines exist for $p=2$; indeed, one may notice from 
the above that there is actually a singular line at $T = \gamma \cdot U$ for any $\gamma \in SL_2(\mathbb{Z})/\Gamma(2)$ --~which is 
not generalisable for arbitrary $p$.

Using the notations introduced in appendix~\ref{app:gamma2}, $\partial_T^3 f^{(1)}_\textsc{het}$ may therefore be written as:
\begin{equation}
	\partial_3^T f^{(1)}_\textsc{het} = f(T) \times \frac{g_{10}(T,U)}{\prod_{\gamma \in SL_2(\mathbb{Z})/\Gamma(2)} \left[\vphantom{\frac{a}{b}} V_\infty(U) (\lambda(U) - \lambda(\gamma \cdot T))\right]}
\end{equation}
where  the modular forms $V_\star (T)$ have a single zero at the corresponding cusp, see eqn.~(\ref{eq:vfunc}), and 
where $g_{10}(T,U)$ is a modular form of weight 10 with respect to $\Gamma(p)_U$; as such, it may be expanded on a basis 
of $\Gamma(p)_U$ modular forms in the variable $U$ as:
\begin{equation}
	g_{10}(T,U) = \sum_{n = 0}^5 a_n(T) X_1^{5-n}(U) X_2^n(U) \, .
\end{equation}

Computing the residue of $\partial_T^3 f^{(1)}_\textsc{het}$ when $U \rightarrow \gamma \cdot T$ is 
made especially easy using equation~\eqref{eq:app_product_lambdas}. One obtains:
\begin{equation}
	\underset{U \rightarrow \gamma \cdot T}{\mathrm{Res}} \partial_T^3 f^{(1)}_\textsc{het} = f(T) \frac{g_{10}(T,\gamma\cdot T)}{\Delta(T) \partial j(T)} J^{-14}(\gamma,T) \overset{!}{=} -\frac{\epsilon_\gamma}{4 \pi^2} \frac{\det^2(\gamma)}{J^{4}(\gamma,T)}
\end{equation}
where equation~\eqref{eq:EFT_residues} has been used to obtain the right-hand side.  Here, $\epsilon_\gamma$ is 1 (resp. -1) 
for singular lines corresponding to vector multiplets (resp. hypermultiplets); these signs reflect  
the respective values of the beta-function coefficient $\beta_\gamma$, which is -4 for a SU(2) without charged hypermultiplet (resp. +4 for a U(1) with two hypermultiplets of charge 1 in absolute value) in the four-dimensional EFT. 

This data provides the necessary information about the behaviour of $g_{10}(T,U)$ under the action of $SL_2(\mathbb{Z})$ on its second variable. 
As moreover $\Gamma(p)$ is normal in $SL_2(\mathbb{Z})$ for any value of $p$, the ring of modular functions of the former is closed under 
the action of the latter. Therefore, $g_{10}(T,\gamma\cdot T)$ is a modular form of weight 10 with respect to $\Gamma(2)$ whose explicit form 
may be computed in terms of $\left\lbrace a_n(T)\right\rbrace$, leading to 5 independent equations allowing one to fix all the $a_n$'s 
(up to an overall multiplicative function of $T$). 

These equations may easily be solved and finally leads to the fully explicit result (see appendix~\ref{app:gamma2}): 
\begin{equation}
	\label{eq:order2_prepotential_correction}
	\partial_T^3 f^{(1)}_\textsc{het} = \frac{i}{995328 \pi} \frac{P_{10,10}(T,U)}{V_0(T) V_1(T) V_\infty(T) \Delta(U) \left[j(T)-j(U)\right]}
\end{equation}
with $P_{10,10}(T,U)$ a modular form of weight $(10,10)$ with respect to $\Gamma(p)_T \times \Gamma(p)_U$ that is given by eqn.~\eqref{eq:P_10_10_explicit}  in appendix~\ref{app:gamma2}. 

Since $T\leftrightarrow U$ exchange is part of the heterotic perturbative duality group, we get immediately the 
expression of $\partial_U^3 f^{(1)}_\textsc{het}$ by exchanging the roles of $T$ and $U$ on the right-hand side of~(\ref{eq:order2_prepotential_correction}).

\section{Hypermultiplets moduli space}
\label{sec:hyper}

In this section we describe the manifold $\mathcal{M}_\textsc{h}$ spanned by the massless scalars in the neutral hypermultiplets of the $\mathcal{N}=2$ low-energy 
four-dimensional theory, which is a quaternionic K\"ahler manifold~\cite{Bagger:1983tt,Cecotti:1988qn}. 

\subsection{Hypermultiplet moduli space in the type IIA description}

From the type IIA perspective, the situation is very different from the usual case of compactifications on Calabi--Yau three-folds. While 
in the latter case $\mathcal{M}_\textsc{h}$ contains the complex structure moduli of the CY$_3$, the scalars from  
the Ramond--Ramond forms and the axio-dilaton (hence receives $g_s$ corrections), in the present case the axio-dilaton lies 
in a vector multiplet and there are no massless fields from the Ramond--Ramond sector~\cite{Israel:2013wwa}. Hence one 
does not expect any correction in the string coupling $g_s$, either perturbative or non-perturbative.

A preliminary analysis of $\mathcal{M}_\textsc{h}$ in type IIA was done in~\cite{Hull:2017llx} and we will summarize now the main results. 
The models can be viewed as orbifolds of a product of a $K3$ Gepner model $\mathcal{G}$ and a $T^2$ by a $\mathbb{Z}_p$ cyclic group acting as a 
non-geometric automorphism of the Gepner model $\mathcal{G}$ and as a shift along the $T^2$. Massless hypermultiplets are obtained from the moduli 
of the type IIA compactification on the $K3$ surface (around the Gepner point $\mathcal{G}$)  invariant under the action of the orbifold. Since 
the orbifold is freely-acting, we cannot get any new hypermultiplet from the twisted sectors of the orbifold.

These considerations indicate that $\mathcal{M}_\textsc{h} \subset \mathcal{M}_\sigma$, {\it i.e.} that this moduli space is a 
subset of the moduli space~(\ref{eq:NLSMmoduli}) of type IIA compactifications on $K3$. An incomplete description of this moduli 
space was obtained in~\cite{Israel:2013wwa}. Consider a $K3$ surface $X$ described by a hypersurface of the form 
\begin{equation}
\label{eq:surf_X}
z_1^{\, p} + f(z_2,z_3,z_4) = 0 
\end{equation}
in a weighted projective space, where $f$ is a quasi-homogeneous polynomial of the appropriate degree. One considers the automorphism 
$\sigma_p : \, z_1 \mapsto e^{2i\pi/p} \, z_1$ and denote by $S(\sigma_p)$ the sub-lattice of the $K3$ lattice $\Gamma_{3,19}$ invariant under 
the action of $\sigma_p$ on the second cohomology. If the K3 surface $X$ is {\it polarized} by the lattice $S(\sigma_p)$ one has an unambiguous 
notion of complex structure on $X$. Then, following recent mathematical results~\cite{Dolgachev2007}, one can determine 
the moduli space $\mathcal{M}_\textsc{cs}^p$ of complex structures on $X$ compatible with the action of $\sigma_p$. 

Consider in the same way $X^\vee$, the Greene--Plesser/Berglund--H\"ubsch mirror of the surface~(\ref{eq:surf_X}), which is a quotient 
of a hypersurface of the form
\begin{equation}
\label{eq:surf_Xvee}
\tilde{z}_1^{\, p} + f^\vee(\tilde{z}_2,\tilde{z}_3,\tilde{z}_4) = 0 \, ,
\end{equation}
in terms of the transpose polynomial $f^\vee$~\cite{Berglund:1991pp}, 
and also admits an order $p$ automorphism acting as $\tilde{\sigma}_p:\, \tilde{z}_1 \mapsto e^{2i\pi/p}\,  \tilde{z}_1$. One can determine 
in the same way the moduli space $\widetilde{\mathcal{M}}_\textsc{cs}^p$ of complex structures on the $S(\tilde{\sigma}_p)$-polarized surface $X^\vee$ that are 
compatible with the action of $\tilde{\sigma}_p$. 

As shown in~\cite{Israel:2013wwa}, the {\it mirrored automorphism} $\widehat{\sigma}_p$ can be viewed as the diagonal action of 
the automorphisms $\sigma_p$ and of $\tilde{\sigma}_p$ on a conformal field theory with target space $X$. Then 
$\mathcal{M}_\textsc{cs}^p \times \widetilde{\mathcal{M}}_\textsc{cs}^p \subset \mathcal{M}_\sigma$ is a sub-manifold 
of the moduli space of conformal field theories on $K3$ surfaces invariant under the action of $\hat{\sigma}_p$, that 
can be viewed as the moduli space of CFTs on $S(\sigma_p)$-polarized $K3$ surfaces invariant under the action of $\hat{\sigma}_p$.

Using the mathematical results known to us, it was not immediate to infer from $\mathcal{M}_\textsc{cs}^p \times \widetilde{\mathcal{M}}_\textsc{cs}^p$ 
the full hypermultiplets moduli space $\mathcal{M}_\textsc{h}$. As we will see below, on the heterotic side of the duality, one can 
get on the nose the exact form of  $\mathcal{M}_\textsc{h}$ by rather standard arguments.

\subsection{Heterotic perspective}

As in type IIA, the heterotic dilaton sits in an $\mathcal{N}=2$ vector multiplet, therefore the hypermultiplet moduli space 
$\mathcal{M}_\textsc{h}$ does not receive corrections either in the string coupling on the heterotic side, hence  can be computed exactly at the perturbative level.

The heterotic description of the non-geometric models~\cite{Gautier:2019qiq} is an order $p$ orbifold acting as an  order $p$ 
$O(\Gamma_{4,20})$ isometry of a heterotic compactification on $T^4$ together with an order $p$ shift along an extra two-torus. The free action of this 
orbifold prevents any (neutral) moduli to arise from the twisted sectors so that the moduli space of the theory should be directly 
inherited of that of the parent theory, that is the heterotic string on $T^4 \times T^2$, and 
the moduli lying in hypermultiplets come from the allowed deformations of the $\Gamma_{4,20}$ lattice associated with the $T^4$ compactification.

Considering the quotient by the automorphism $\hat{\sigma}_p$ only makes sense for lattices $\Gamma_{4,20}$ which admit 
$\hat{\sigma}_p$ as a symmetry, which means that the moduli space we are looking for may be interpreted as the space of 
deformations of such lattices. Phrased differently, the local form of $\mathcal{M}_H$ may be accessed by picking such a particular 
lattice and studying its deformations compatible with  $\hat{\sigma}_p$-invariance. 

At this stage, one may emphasise the peculiarity of the $p=2$ model: first of all, one may notice that 
equation~(\ref{eq:lemma_diag}) implies that the matrix $M_2$ associated with the action of $\hat{\sigma}_2$ 
is simply $M_2 = - \mathbb{I}_{24}$; as any lattice admits 
this order two symmetry, all the deformations of $\Gamma_{4,20}$ are still allowed in the orbifolded theory, which 
is clearly not the case for any other admissible value of $p$. Therefore, in the $p=2$ case, the hypermultiplets 
moduli space is directly given by the $T^4$ moduli space of the six-dimensional compactification:

\begin{equation}
\mathcal{M}_\textsc{h}^{p=2} \cong \moduli{O(\Gamma_{4,20})}{O(4,20)}{O(4) \times O(20)}
\end{equation}

Let us then consider the other cases where $p>2$. The deformations of the Narain lattice $\Gamma_{4,20}$ in the parent theory correspond locally to choices of embedding 
of this lattice into $\mathbb{R}^{4,20}$, that is to fixing a space-like 4-dimensional plane $\Pi_L(\Gamma_{4,20})$ in the ambient space of 
$\Gamma_{4,20}$ (or equivalently to fixing a time-like 20-dimensional plane $\Pi_R(\Gamma_{4,20})$). 

Given that, in the type IIA description, the automorphism $\sigma_p$ acts by definition on the holomorphic two-form $\omega$ as $\omega \mapsto \zeta_p 
\omega$ 
and that $\int \omega\wedge \bar \omega >0$, there exists a space-like eigenspace associated with the eigenvalue $\zeta_p$ of the automorphism. Further, on the heterotic 
side, $\mathcal{N}=2$ space-time supersymmetry of the asymmetric orbifold~\cite{Gautier:2019qiq} 
indicates that there must exist a basis of $\Pi_L(\Gamma_{4,20}) \otimes \mathbb{C}$ in which 
$M_p = \mathrm{diag}(\zeta_p \mathbb{I}_2, \zeta_p^{-1} \mathbb{I}_2)$.

Then, given the diagonal action of $M_p$ onto $\Pi_L(\Gamma_{4,20}) \otimes \mathbb{C}$, the only freedom left amounts to choosing which directions 
correspond to left-movers in the eigenspace of $\hat{\sigma}_p$ corresponding to, say, $\zeta_p$. Moreover, 
equation~(\ref{eq:lemma_diag}) states that the eigenspace of any eigenvalue of $\hat{\sigma}_p$ has 
dimension $24/\varphi(p)$, so that the moduli lying in hypermultiplets may be understood as arising from the freedom of 
choice of a space-like 2-dimensional complex plane into a $24/\varphi(p)$-dimensional complex space.  Therefore, $\mathcal{M}_H$ 
may be locally understood as a Grassmannian space of complex spaces and has the local form:

\begin{equation}
\mathcal{T}^{p \neq 2}_\textsc{h} \cong \teich{SU\left(2,\frac{24}{\varphi(p)}-2\right)}{S\left[U(2) \times U\left(\frac{24}{\varphi(p)}-2\right)\right]}.
\label{eq:teich_hyper}
\end{equation}

The global form of $\mathcal{M}_\textsc{h}$ is then obtained by identifying the corresponding duality group; this is done by noticing that this theory 
inherits its dualities from the mother toroidal theory. Indeed, as the orbifold procedure keeps only $\hat{\sigma}_p$-invariant states, any element 
$g$ of the duality group must commute with the induced action $\hat{\sigma}_p^\star$ of the automorphism on the states of the theory. Then, any 
element $g$ of the duality group of the original theory, that is $O(\Gamma_{4,20})$, satisfying such a condition must belong also to the duality 
group of the resulting theory. Furthermore, a little bit of thought also shows that no other duality element may be present here, as 
$O(\Gamma_{4,20})$ already includes all acceptable duality relations inside a given sector; 
indeed, new elements would necessarily mix states from different sectors, which is not possible as they would have different conformal 
weights due to the free action of the orbifolds we are interested in. In summary, defining
\begin{equation}
\hat{O}_p := \left\lbrace\gamma \in O(\Gamma_{4,20}) \middle| \gamma \circ \hat{\sigma}_p^\star = \hat{\sigma}_p^\star \circ \gamma \right\rbrace ,
\end{equation}
the full moduli space spanned by scalars in hypermultiplets reads
\begin{equation}
\label{eq:hyper_moduli_space}
\mathcal{M}^{p \neq 2}_\textsc{h} \cong 
\moduli{\hat{O}_p}{SU\left(2,\frac{24}{\varphi(p)}-2\right)}{S\left[U(2) \times U\left(\frac{24}{\varphi(p)}-2\right)\right]}.
\end{equation}

We have checked that the above analysis is also compatible with the BPS indices obtained in~\cite{Gautier:2019qiq}, from which one can 
infer in particular the difference $n_V - n_H$ between the number of massless vector and hypermultiplets. It may also be noted that 
similar types of hypermultiplets moduli spaces have already been considered in the  literature, 
see \textit{e.g.}~\cite{Ferrara:1988fr}, where it was noticed in particular that~(\ref{eq:teich_hyper}) was indeed a quaternionic K\"ahler manifold.

This hypermultiplets moduli space does not receive by construction corrections in the string coupling $g_s$. Crucially, one can 
argue that it does not receive $\alpha'$ corrections as well. The moduli space is derived in the heterotic 
description from an exact toroidal CFT on the worldsheet and, 
due to the freely-acting nature of the orbifold, there are no moduli from the twisted sectors. Besides this, as was recalled in 
section~\ref{sec:review}, the heterotic models at hand do not admit any non-abelian gauge bundle hence there are no 
small-instanton singularities anywhere in the moduli space.

\section{Conclusions}
\label{sec:concl}

In this work we have derived the moduli space of $\mathcal{N}=2$ four-dimensional compactifications on non-geometric backgrounds, using both 
their description in the type IIA duality frame as non-geometric Calabi-Yau backgrounds~\cite{Hull:2017llx} and their description in the heterotic 
frame as asymmetric and freely-acting toroidal orbifolds~\cite{Gautier:2019qiq}.

We have first analyzed the vector multiplets moduli space, which receives corrections in the string coupling 
both in type IIA and in heterotic frames. While, 
as we have shown, there are only non-perturbative corrections to the prepotential in the type IIA variables, the heterotic prepotential 
receives both one-loop and non-perturbative corrections w.r.t. the heterotic dilaton. 

Thanks to an analysis of the perturbative duality group acting on the heterotic vector multiplets moduli space --~or at least of a subgroup of it~-- we have 
shown how to obtain  an explicit expression of the third derivative of the one-loop prepotential, using that the latter 
is a modular form in the  $T$ and $U$ variables ({\it i.e.} the moduli of the heterotic $T^2$) with respect to a $\Gamma (p) \times \Gamma (p)$ 
subgroup of the duality group. We have given explicitely the result for mirrored automorphisms of order $p=2$ and explained how 
to generalize this result to mirrored automorphisms of arbitrary order $p>2$. It would be interesting to obtain explicit results in those 
cases as well. 

Finally we have studied the hypermultiplets moduli space, which is exact in the string coupling constant on both sides of the duality. 
While obtaining the hypermultiplets moduli space from a type IIA perspective is not trivial (see~\cite{Hull:2017llx} for a discussion), the 
heterotic description of the models as asymmetric toroidal orbifold allowed us to get an exact description of these 
moduli spaces both in $\alpha'$ and in $g_s$. 

For the models studied in this paper the situation is  in some way the opposite of what was found for dualities  
between type IIA on Calabi-Yau threefolds and heterotic on $K3 \times T^2$. 
In the latter case, there exists a duality frame in which the vector moduli space can be computed classically, while the hypermultiplets 
receives corrections in both frames (either in $g_s$ or in $\alpha'$) that are not yet fully understood. In the present case, while the hypermultiplet 
moduli space is exact (in a rather mundane way) the vector multiplets moduli space receives $g_s$ corrections in any duality frame.

Building on the results of this paper, it would be very interesting to analyse explicitely the perturbative heterotic corrections to the prepotential from a 
type IIA perspective, as they are expected to correspond to NS5-brane instantons wrapping 'cycles' of non-geometric backgrounds, which are, by 
definition, quite tricky to study directly. The asymptotic expansion of the results obtained in the present work may allow to understand whether a 
semi-classical analysis of such instantons exists in non-geometric backgrounds, thereby providing insights into their quantum (non-)geometry.

Also, one may wonder whether comparing the perturbative expansions of the vector multiplets prepotential on the type IIA side and on the heterotic side may 
allow to guess its exact non-perturbative form.

\section*{Acknowledgements}

We acknowledge the involvement of Chris Hull at the first stages of this project. We would like to thank  Chris Hull, Boris Pioline and Alessandra Sarti for numerous discussions and suggestions.

\section{Appendix}

\subsection{\texorpdfstring{$\vartheta$}{Theta}-functions and their properties}

In the following, we briefly remind the definition of the Jacobi $\vartheta$-functions used in~\cite{Gautier:2019qiq} as well as in this paper. We then define the Jacobi $\vartheta$ function with characteristic as:
\begin{equation}
  \vartheta\jtheta{\alpha}{\beta} \left(\tau \middle| v \right) := \sum_{n \in \mathbb{Z}} q^{\frac{1}{2} \left(n+\frac{\alpha}{2}\right)^2} e^{2 i \pi \left(n + \frac{\alpha}{2}\right)\left(v+\frac{\beta}{2}\right)} ,
\end{equation}
where $\alpha, \beta \in \mathbb{R}$ and where $q := \exp(2 i \pi \tau)$.

As recalled in equation~\eqref{eq:lemma_diag}, the eigenvalues of the $\Gamma_{4,20}$ automorphism used to generate the $\mathbb{Z}_p$ orbifold are all primitive $p$-th roots of unity, each with the same multiplicity. Consequently, when computing one-loop quantities, the product
\begin{equation}
	\Theta_g^{(p)}(\tau) := \prod_{\substack{t = 1\\(t,p) = 1}}^{p-1} \vartheta\jtheta{1}{1-2 g t/p}(\tau)
\end{equation}
usually appears through the contribution of the untwisted sector. Here, $(t,p)$ is a shorthand notation for $\gcd(t,p)$. It will then be helpful, at least for numerical computations, to notice that $\Theta_g^{(p)}$ may be rewritten as

\begin{equation}
	\label{eq:theta_to_eta}
	\Theta_g^{(p)}(\tau) = e^{\frac{i\pi}{2}(1-g) \varphi(p)} \eta^{\varphi(p)}(\tau) \left[\Phi_{x_g}(1) \prod_{d | x_g} \eta(d \tau)^{2 \mu\left(\frac{x_g}{d}\right)} \right]^{\frac{\varphi(p)}{\varphi(x_g)}}
\end{equation}
with $x_g := \frac{p}{(g,p)}$, $\varphi$ the Euler totient function as before, $\mu$ the M\"obius function and $\Phi_{n}(x)$ the $n$-th cyclotomic polynomial.

\subsection{\texorpdfstring{$\Gamma(2)$}{Gamma(2)} modular forms}
\label{app:gamma2}

The ring of modular forms of $\Gamma(2)$ is knoen to be isomorphic to the ring of modular forms of $\Gamma_0(4)$ (the set of $SL_2(\mathbb{Z}$ elements with 
vanishing lower-left component modulo 4), see \textit{e.g.}~\cite{Angelantonj:2013eja} for details.

A basis of the modular ring of $\Gamma(2)$ is given by  the modular forms $X_1$ and $X_2$, defined as:
\begin{equation}
X_1(\tau) := E_2(\tau) - 2 E_2(2 \tau) \qquad X_2(\tau) := E_2\left(\frac{1}{2} \tau\right) - 4 E_2(2 \tau) \, ,
\end{equation}
with $E_2$ the Eisenstein series of weight 2 defined as:
\begin{equation}
E_2(\tau) := 1-24 \sum_{n=1}^\infty \sigma_1(n) q^n \, ,
\end{equation}
$\sigma_1(n)$ being the sum of the divisors of $n$, which is not a modular form.

Modular forms of weight $2k$ of $\Gamma(2)$ have exactly $k$ zeroes; it will therefore be useful to also define:
\begin{equation}
\label{eq:vfunc}
\begin{split}
V_0(\tau) & := \frac{6 X_1(\tau)-X_2(\tau)}{48} \\
V_1(\tau) & := - \frac{X_2(\tau)}{48} \\
V_\infty(\tau) & := \frac{3 X_1-X_2(\tau)}{24}
\end{split}
\end{equation}
so that $V_s$ vanishes as $\sqrt{q} + \mathcal{O}(q)$ around the cusp $s$, with $s=0, 1, \infty$ and $q:= \exp(2 i \pi \tau)$. 

The Hauptmodul for the congruence subgroup $\Gamma(2)$ is the well-known $\lambda$ function, expressed in terms of the above as follows:
\begin{equation}
\lambda(\tau) := 16 \frac{V_1(\tau)}{V_\infty(\tau)} \, .
\end{equation}
A useful relation, allowing one to compactly write the results of section~\ref{subsec:order2_explicit}, is:
\begin{equation}
\label{eq:app_product_lambdas}
\prod_{\gamma \in SL_2(\mathbb{Z})/\Gamma(2)} \left[\vphantom{\frac{a}{b}} \lambda(U)-\lambda(\gamma\cdot T)\right] = \frac{\Delta(U)}{V_\infty^6(U)} \left[\vphantom{\frac{a}{b}}j(U)-j(T)\right]
\end{equation}
with $j$ the Klein $j$ function, that is the $SL_2(\mathbb{Z})$ $j$-invariant, and $\Delta(U)$ the cusp form of $SL_2(\mathbb{Z})$.

This peculiar occurrence of modular functions of $SL_2(\mathbb{Z})$ instead of just $\Gamma(2)$ is proper to the $p=2$ case. 
It is linked to the fact that the singular lines are located at $T = \gamma \cdot U$ for all $\gamma$ in the 
coset $SL_2(\mathbb{Z})/\Gamma(2)$, so that in the end one has a singular line whenever 
$T = g \cdot U$ for any $g$ in $SL_2(\mathbb{Z})$ (even though every such $g$ would not lead to physically equivalent configurations).

The modular form $P_{10,10}$ appearing in equation~\eqref{eq:order2_prepotential_correction} reads, in terms of the above modular forms,
\begin{align}
	\label{eq:P_10_10_explicit}
	\begin{split}
		P_{10,10}(T,U) & = -32 X_1(T)^5 X_2(U)^5+288 X_1(T)^5 X_1(U) X_2(U)^4 \\
		& -576 X_1(T)^5 X_1(U)^2 X_2(U)^3+24 X_1(T)^4 X_2(T) X_2(U)^5 \\
		& -8 X_1(T)^4 X_2(T) X_1(U) X_2(U)^4-2304 X_1(T)^4 X_2(T) X_1(U)^2 X_2(U)^3 \\
		& +12672 X_1(T)^4 X_2(T) X_1(U)^3 X_2(U)^2+20736 X_1(T)^4
		X_2(T) X_1(U)^5 \\
		& -25920 X_1(T)^4 X_2(T) X_1(U)^4 X_2(U)-4 X_1(T)^3 X_2(T)^2 X_2(U)^5 \\
		& -192 X_1(T)^3 X_2(T)^2 X_1(U)
		X_2(U)^4+2848 X_1(T)^3 X_2(T)^2 X_1(U)^2 X_2(U)^3 \\
		& -20736 X_1(T)^3 X_2(T)^2 X_1(U)^5-13248 X_1(T)^3 X_2(T)^2 X_1(U)^3
		X_2(U)^2 \\
		& +26496 X_1(T)^3 X_2(T)^2 X_1(U)^4 X_2(U)+88 X_1(T)^2 X_2(T)^3 X_1(U) X_2(U)^4 \\
		& +7488 X_1(T)^2 X_2(T)^3 X_1(U)^5-1104
		X_1(T)^2 X_2(T)^3 X_1(U)^2 X_2(U)^3 \\
		& +4960 X_1(T)^2 X_2(T)^3 X_1(U)^3 X_2(U)^2-9792 X_1(T)^2 X_2(T)^3 X_1(U)^4 X_2(U) \\
		& -1152
		X_1(T) X_2(T)^4 X_1(U)^5-15 X_1(T) X_2(T)^4 X_1(U) X_2(U)^4 \\
		& +184 X_1(T) X_2(T)^4 X_1(U)^2 X_2(U)^3-816 X_1(T) X_2(T)^4
		X_1(U)^3 X_2(U)^2 \\
		& +1576 X_1(T) X_2(T)^4 X_1(U)^4 X_2(U)+64 X_2(T)^5 X_1(U)^5 \\
		& +X_2(T)^5 X_1(U) X_2(U)^4-12 X_2(T)^5
		X_1(U)^2 X_2(U)^3 \\
		& +52 X_2(T)^5 X_1(U)^3 X_2(U)^2-96 X_2(T)^5 X_1(U)^4 X_2(U)
	\end{split}
\end{align}

\bibliography{biblio_moduli}

\providecommand{\href}[2]{#2}\begingroup\raggedright\begin{thebibliography}{10}

\bibitem{Aspinwall:2000fd}
P.~S. Aspinwall, {\it {Compactification, geometry and duality: N=2}},  in {\em
  {Strings, branes and gravity. Proceedings, Theoretical Advanced Study
  Institute, TASI'99, Boulder, USA, May 31-June 25, 1999}}, pp.~723--805, 2000.
\newblock \href{http://xxx.lanl.gov/abs/hep-th/0001001}{{\tt hep-th/0001001}}.

\bibitem{Kachru:1995wm}
S.~Kachru and C.~Vafa, {\it {Exact results for N=2 compactifications of
  heterotic strings}},  {\em Nucl. Phys.} {\bf B450} (1995) 69--89,
  [\href{http://xxx.lanl.gov/abs/hep-th/9505105}{{\tt hep-th/9505105}}].

\bibitem{Ferrara:1995yx}
S.~Ferrara, J.~A. Harvey, A.~Strominger, and C.~Vafa, {\it {Second quantized
  mirror symmetry}},  {\em Phys. Lett.} {\bf B361} (1995) 59--65,
  [\href{http://xxx.lanl.gov/abs/hep-th/9505162}{{\tt hep-th/9505162}}].

\bibitem{Vafa:1995gm}
C.~Vafa and E.~Witten, {\it {Dual string pairs with N=1 and N=2 supersymmetry
  in four-dimensions}},  {\em Nucl. Phys. Proc. Suppl.} {\bf 46} (1996)
  225--247, [\href{http://xxx.lanl.gov/abs/hep-th/9507050}{{\tt
  hep-th/9507050}}]. [,225(1995)].

\bibitem{Hull:1994ys}
C.~M. Hull and P.~K. Townsend, {\it {Unity of superstring dualities}},  {\em
  Nucl. Phys.} {\bf B438} (1995) 109--137,
  [\href{http://xxx.lanl.gov/abs/hep-th/9410167}{{\tt hep-th/9410167}}].

\bibitem{Aspinwall:1995fw}
P.~S. Aspinwall, {\it {Some relationships between dualities in string theory}},
   {\em Nucl. Phys. Proc. Suppl.} {\bf 46} (1996) 30--38,
  [\href{http://xxx.lanl.gov/abs/hep-th/9508154}{{\tt hep-th/9508154}}].

\bibitem{Ferrara:1989nm}
S.~Ferrara and C.~Kounnas, {\it {Extended Supersymmetry in Four-dimensional
  Type {II} Strings}},  {\em Nucl. Phys.} {\bf B328} (1989) 406--438.

\bibitem{Schwarz:1995bj}
J.~H. Schwarz and A.~Sen, {\it {Type IIA dual of the six-dimensional CHL
  compactification}},  {\em Phys. Lett.} {\bf B357} (1995) 323--328,
  [\href{http://xxx.lanl.gov/abs/hep-th/9507027}{{\tt hep-th/9507027}}].

\bibitem{Chaudhuri:1995fk}
S.~Chaudhuri, G.~Hockney, and J.~D. Lykken, {\it {Maximally supersymmetric
  string theories in D < 10}},  {\em Phys. Rev. Lett.} {\bf 75} (1995)
  2264--2267, [\href{http://xxx.lanl.gov/abs/hep-th/9505054}{{\tt
  hep-th/9505054}}].

\bibitem{Datta:2015hza}
S.~Datta, J.~R. David, and D.~Lust, {\it {Heterotic string on the CHL orbifold
  of K3}},  {\em JHEP} {\bf 02} (2016) 056,
  [\href{http://xxx.lanl.gov/abs/1510.0542}{{\tt arXiv:1510.0542}}].

\bibitem{Gautier:2019qiq}
Y.~Gautier, C.~M. Hull, and D.~Israel, {\it {Heterotic/type II Duality and
  Non-Geometric Compactifications}},  {\em JHEP} {\bf 10} (2019) 214,
  [\href{http://xxx.lanl.gov/abs/1906.0216}{{\tt arXiv:1906.0216}}].

\bibitem{Israel:2013wwa}
D.~Israel and V.~Thiery, {\it {Asymmetric Gepner models in type II}},  {\em
  JHEP} {\bf 02} (2014) 011, [\href{http://xxx.lanl.gov/abs/1310.4116}{{\tt
  arXiv:1310.4116}}].

\bibitem{Israel:2015efa}
D.~Israel, {\it {Nongeometric Calabi-Yau compactifications and fractional
  mirror symmetry}},  {\em Phys. Rev.} {\bf D91} (2015) 066005,
  [\href{http://xxx.lanl.gov/abs/1503.0155}{{\tt arXiv:1503.0155}}]. [Erratum:
  Phys. Rev.D91,no.12,129902(2015)].

\bibitem{Hull:2017llx}
C.~Hull, D.~Israel, and A.~Sarti, {\it {Non-geometric Calabi-Yau Backgrounds
  and K3 automorphisms}},  {\em JHEP} {\bf 11} (2017) 084,
  [\href{http://xxx.lanl.gov/abs/1710.0085}{{\tt arXiv:1710.0085}}].

\bibitem{Hull:2004in}
C.~M. Hull, {\it {A Geometry for non-geometric string backgrounds}},  {\em
  JHEP} {\bf 10} (2005) 065,
  [\href{http://xxx.lanl.gov/abs/hep-th/0406102}{{\tt hep-th/0406102}}].

\bibitem{Kaplunovsky:1995tm}
V.~Kaplunovsky, J.~Louis, and S.~Theisen, {\it {Aspects of duality in N=2
  string vacua}},  {\em Phys. Lett.} {\bf B357} (1995) 71--75,
  [\href{http://xxx.lanl.gov/abs/hep-th/9506110}{{\tt hep-th/9506110}}].

\bibitem{Antoniadis:1995zn}
I.~Antoniadis, E.~Gava, K.~S. Narain, and T.~R. Taylor, {\it {N=2 type II
  heterotic duality and higher derivative F terms}},  {\em Nucl. Phys.} {\bf
  B455} (1995), no.~1-2 109--130,
  [\href{http://xxx.lanl.gov/abs/hep-th/9507115}{{\tt hep-th/9507115}}].

\bibitem{Harvey:1995fq}
J.~A. Harvey and G.~W. Moore, {\it {Algebras, BPS states, and strings}},  {\em
  Nucl. Phys.} {\bf B463} (1996) 315--368,
  [\href{http://xxx.lanl.gov/abs/hep-th/9510182}{{\tt hep-th/9510182}}].

\bibitem{Alexandrov:2013yva}
S.~Alexandrov, J.~Manschot, D.~Persson, and B.~Pioline, {\it {Quantum
  hypermultiplet moduli spaces in N=2 string vacua: a review}},  {\em Proc.
  Symp. Pure Math.} {\bf 90} (2015) 181--212,
  [\href{http://xxx.lanl.gov/abs/1304.0766}{{\tt arXiv:1304.0766}}].

\bibitem{Antoniadis:1995ct}
I.~Antoniadis, S.~Ferrara, E.~Gava, K.~S. Narain, and T.~R. Taylor, {\it
  {Perturbative prepotential and monodromies in N=2 heterotic superstring}},
  {\em Nucl. Phys.} {\bf B447} (1995) 35--61,
  [\href{http://xxx.lanl.gov/abs/hep-th/9504034}{{\tt hep-th/9504034}}].

\bibitem{deWit:1995dmj}
B.~de~Wit, V.~Kaplunovsky, J.~Louis, and D.~Lust, {\it {Perturbative couplings
  of vector multiplets in N=2 heterotic string vacua}},  {\em Nucl. Phys.} {\bf
  B451} (1995) 53--95, [\href{http://xxx.lanl.gov/abs/hep-th/9504006}{{\tt
  hep-th/9504006}}].

\bibitem{Seiberg:1988pf}
N.~Seiberg, {\it {Observations on the Moduli Space of Superconformal Field
  Theories}},  {\em Nucl. Phys.} {\bf B303} (1988) 286--304.

\bibitem{Aspinwall:1994rg}
P.~S. Aspinwall and D.~R. Morrison, {\it {String theory on K3 surfaces}},
  \href{http://xxx.lanl.gov/abs/hep-th/9404151}{{\tt hep-th/9404151}}.

\bibitem{mirror0}
M.~Artebani, {S. Boissi\`ere}, and A.~Sarti, {\it {The
  Berglund-H\"ubsch-Chiodo-Ruan mirror symmetry for K3 surfaces}},  {\em
  Journal de Mathematiques Pures et Appliquees} {\bf 102} (2014), no.~4 758 --
  781.

\bibitem{Comparin:2012ps}
P.~Comparin, C.~Lyons, N.~Priddis, and R.~Suggs, {\it {The mirror symmetry of
  K3 surfaces with non-symplectic automorphisms of prime order}},  {\em Adv.
  Theor. Math. Phys.} {\bf 18} (2014), no.~6 1335--1368,
  [\href{http://xxx.lanl.gov/abs/1211.2172}{{\tt arXiv:1211.2172}}].

\bibitem{mirror1}
P.~Comparin and N.~Priddis, {\it {BHK mirror symmetry for K3 surfaces with
  non-symplectic automorphism}},  \href{http://xxx.lanl.gov/abs/1704.0035}{{\tt
  arXiv:1704.0035}}.

\bibitem{mirror2}
C.~Bott, P.~Comparin, and N.~Priddis, {\it {Mirror symmetry for K3 surfaces}},
  \href{http://xxx.lanl.gov/abs/1901.0937}{{\tt arXiv:1901.0937}}.

\bibitem{Greene:1990ud}
B.~R. Greene and M.~R. Plesser, {\it {Duality in {Calabi-Yau} Moduli Space}},
  {\em Nucl. Phys.} {\bf B338} (1990) 15--37.

\bibitem{Berglund:1991pp}
P.~Berglund and T.~Hubsch, {\it {A Generalized construction of mirror
  manifolds}},  {\em Nucl. Phys.} {\bf B393} (1993) 377--391,
  [\href{http://xxx.lanl.gov/abs/hep-th/9201014}{{\tt hep-th/9201014}}].
  [,327(1991); AMS/IP Stud. Adv. Math.9,327(1998)].

\bibitem{Dabholkar:2002sy}
A.~Dabholkar and C.~Hull, {\it {Duality twists, orbifolds, and fluxes}},  {\em
  JHEP} {\bf 09} (2003) 054,
  [\href{http://xxx.lanl.gov/abs/hep-th/0210209}{{\tt hep-th/0210209}}].

\bibitem{Dabholkar:2005ve}
A.~Dabholkar and C.~Hull, {\it {Generalised T-duality and non-geometric
  backgrounds}},  {\em JHEP} {\bf 05} (2006) 009,
  [\href{http://xxx.lanl.gov/abs/hep-th/0512005}{{\tt hep-th/0512005}}].

\bibitem{Narain:1986qm}
K.~S. Narain, M.~H. Sarmadi, and C.~Vafa, {\it {Asymmetric Orbifolds}},  {\em
  Nucl. Phys.} {\bf B288} (1987) 551.

\bibitem{deWit:1984wbb}
B.~de~Wit and A.~Van~Proeyen, {\it {Potentials and Symmetries of General Gauged
  N=2 Supergravity: Yang-Mills Models}},  {\em Nucl. Phys.} {\bf B245} (1984)
  89--117.

\bibitem{Antoniadis:1992pm}
I.~Antoniadis, E.~Gava, K.~S. Narain, and T.~R. Taylor, {\it {Superstring
  threshold corrections to Yukawa couplings}},  {\em Nucl. Phys.} {\bf B407}
  (1993) 706--724, [\href{http://xxx.lanl.gov/abs/hep-th/9212045}{{\tt
  hep-th/9212045}}].

\bibitem{Forger:1997tu}
K.~Forger and S.~Stieberger, {\it {String amplitudes and N=2, d = 4
  prepotential in heterotic K3 x T**2 compactifications}},  {\em Nucl. Phys.}
  {\bf B514} (1998) 135--160,
  [\href{http://xxx.lanl.gov/abs/hep-th/9709004}{{\tt hep-th/9709004}}].

\bibitem{Angelantonj:2011br}
C.~Angelantonj, I.~Florakis, and B.~Pioline, {\it {A new look at one-loop
  integrals in string theory}},  {\em Commun. Num. Theor. Phys.} {\bf 6} (2012)
  159--201, [\href{http://xxx.lanl.gov/abs/1110.5318}{{\tt arXiv:1110.5318}}].

\bibitem{Angelantonj:2012gw}
C.~Angelantonj, I.~Florakis, and B.~Pioline, {\it {One-Loop BPS amplitudes as
  BPS-state sums}},  {\em JHEP} {\bf 06} (2012) 070,
  [\href{http://xxx.lanl.gov/abs/1203.0566}{{\tt arXiv:1203.0566}}].

\bibitem{Angelantonj:2013eja}
C.~Angelantonj, I.~Florakis, and B.~Pioline, {\it {Rankin-Selberg methods for
  closed strings on orbifolds}},  {\em JHEP} {\bf 07} (2013) 181,
  [\href{http://xxx.lanl.gov/abs/1304.4271}{{\tt arXiv:1304.4271}}].

\bibitem{Angelantonj:2015rxa}
C.~Angelantonj, I.~Florakis, and B.~Pioline, {\it {Threshold corrections,
  generalised prepotentials and Eichler integrals}},  {\em Nucl. Phys.} {\bf
  B897} (2015) 781--820, [\href{http://xxx.lanl.gov/abs/1502.0000}{{\tt
  arXiv:1502.0000}}].

\bibitem{Cecotti:1992qh}
S.~Cecotti, P.~Fendley, K.~A. Intriligator, and C.~Vafa, {\it {A New
  supersymmetric index}},  {\em Nucl. Phys.} {\bf B386} (1992) 405--452,
  [\href{http://xxx.lanl.gov/abs/hep-th/9204102}{{\tt hep-th/9204102}}].

\bibitem{Kiritsis:1994ta}
E.~Kiritsis and C.~Kounnas, {\it {Infrared regularization of superstring theory
  and the one loop calculation of coupling constants}},  {\em Nucl. Phys.} {\bf
  B442} (1995) 472--493, [\href{http://xxx.lanl.gov/abs/hep-th/9501020}{{\tt
  hep-th/9501020}}].

\bibitem{Kaplunovsky:1995jw}
V.~Kaplunovsky and J.~Louis, {\it {On Gauge couplings in string theory}},  {\em
  Nucl. Phys.} {\bf B444} (1995) 191--244,
  [\href{http://xxx.lanl.gov/abs/hep-th/9502077}{{\tt hep-th/9502077}}].

\bibitem{Persson:2015jka}
D.~Persson and R.~Volpato, {\it {Fricke S-duality in CHL models}},  {\em JHEP}
  {\bf 12} (2015) 156, [\href{http://xxx.lanl.gov/abs/1504.0726}{{\tt
  arXiv:1504.0726}}].

\bibitem{Kiritsis:1996xd}
E.~Kiritsis, C.~Kounnas, P.~M. Petropoulos, and J.~Rizos, {\it {Solving the
  decompactification problem in string theory}},  {\em Phys. Lett.} {\bf B385}
  (1996) 87--95, [\href{http://xxx.lanl.gov/abs/hep-th/9606087}{{\tt
  hep-th/9606087}}].

\bibitem{Aspinwall:1999ii}
P.~S. Aspinwall and M.~Plesser, {\it {T duality can fail}},  {\em JHEP} {\bf
  08} (1999) 001, [\href{http://xxx.lanl.gov/abs/hep-th/9905036}{{\tt
  hep-th/9905036}}].

\bibitem{diamond2006first}
F.~Diamond and J.~Shurman, {\em A First Course in Modular Forms}.
\newblock Graduate Texts in Mathematics. Springer New York, 2006.

\bibitem{Bagger:1983tt}
J.~Bagger and E.~Witten, {\it {Matter Couplings in N=2 Supergravity}},  {\em
  Nucl. Phys.} {\bf B222} (1983) 1--10.

\bibitem{Cecotti:1988qn}
S.~Cecotti, S.~Ferrara, and L.~Girardello, {\it {Geometry of Type II
  Superstrings and the Moduli of Superconformal Field Theories}},  {\em Int. J.
  Mod. Phys.} {\bf A4} (1989) 2475.

\bibitem{Dolgachev2007}
I.~V. Dolgachev and S.~Kond{\={o}}, {\em {Moduli of K3 Surfaces and Complex
  Ball Quotients}}, pp.~43--100.
\newblock {Birkh{\"a}user Basel}, Basel, 2007.

\bibitem{Ferrara:1988fr}
S.~Ferrara and M.~Porrati, {\it {The Manifolds of Scalar Background Fields in
  $Z(N$) Orbifolds}},  {\em Phys. Lett.} {\bf B216} (1989) 289--296.

\end{thebibliography}\endgroup

\end{document}